\documentclass[journal]{IEEEtran}
\usepackage{verbatim}
\usepackage[utf8x]{inputenc}
\usepackage{epstopdf}
\usepackage[pdftex]{graphicx}
\usepackage{graphics}
\usepackage{tabularx}
\usepackage{booktabs}
\usepackage[table]{xcolor}
\usepackage{multirow}
\usepackage{siunitx}
\usepackage{tabu}
\usepackage{booktabs} \usepackage{ctable}
\usepackage[cmex10]{amsmath}
\usepackage{amssymb}
\usepackage{cite}
\usepackage{textcomp}
\usepackage{array}
\newcolumntype{C}[1]{>{\centering\let\newline\\\arraybackslash\hspace{0pt}}m{#1}}

\graphicspath{{./Figures/}}

\ifCLASSOPTIONcompsoc
\usepackage[tight,normalsize,sf,SF]{subfigure}
\else
\usepackage[tight,footnotesize]{subfigure}
\fi

\begin{document}
\title{Multiband Spectrum Access: Great Promises for Future Cognitive Radio Networks} 
\author{Ghaith Hattab,~\IEEEmembership{Student Member,~IEEE,} Mohammed Ibnkahla,~\IEEEmembership{Member,~IEEE}
\thanks{Manuscript received.}%
\thanks{G. Hattab and M. Ibnkahla are with the department of Electrical and Computer Engineering at Queen's University, Kingston, ON, Canada (email: g.hattab@queensu.ca, and mohamed.ibnkahla@queensu.ca).}}

\maketitle

\begin{abstract}
Cognitive radio has been widely considered as one of the prominent solutions to tackle the spectrum scarcity. While the majority of existing research has focused on single-band cognitive radio, multiband cognitive radio represents great promises towards implementing efficient cognitive networks compared to single-based networks. Multiband cognitive radio networks (MB-CRNs) are expected to significantly enhance the network's throughput and provide better channel maintenance by reducing handoff frequency. Nevertheless, the wideband front-end and the multiband spectrum access impose a number of challenges yet to overcome. This paper provides an in-depth analysis on the recent advancements in multiband spectrum sensing techniques, their limitations, and possible future directions to improve them. We study cooperative communications for MB-CRNs to tackle a fundamental limit on diversity and sampling. We also investigate several limits and tradeoffs of various design parameters for MB-CRNs. In addition, we explore the key MB-CRNs performance metrics that differ from the conventional metrics used for single-band based networks.
\end{abstract}


\bstctlcite{IEEEexample:BSTcontrol}

\section{Introduction}
\IEEEPARstart{C}{ognitive} radio (CR) has been widely considered as one of the prominent solutions to spectrum scarcity that arises due to the abundance of wireless technologies, the ever-increasing demand for higher data rates, the regulated static spectrum allocation, and the underutilization of the entire spectrum at a certain time, frequency, or space \cite{Mitola1, FCC2, Haykin}. Cognitive radio networks (CRNs) are expected to efficiently utilize the unused spectrum by enabling unlicensed users (also known as secondary users (SUs)) to opportunistically access or share the licensed bands with the licensed users (also known as primary users (PUs)).

There are certain requirements on the CRNs that distinguish them from the conventional wireless communication networks. In particular, CRNs must tackle the coexistence of SUs with the PUs. Such coexistence could be handled in three different paradigms. For instance, in \emph{interweave} paradigm, the SU must not access a band when the PU transmits over it. However, in \emph{underlay} and \emph{overlay} paradigms, the SU may concurrently access a band with the PU. In the former, an interference limit must be considered by the SU to protect the PUs, and in the latter, the SU must have a large amount of side information about the PU network (e.g., channel conditions, PU encoding techniques, etc.) to coexist with the PU with minimum interference \cite[ch.~2]{Biglieri}. These different access paradigms require reliable \emph{spectrum awareness} to obtain information about the spectrum status \cite{Ibnkahla1}.

Spectrum awareness is classified into \emph{passive}-based and \emph{active}-based awareness. In the former, the SU learns about the spectrum availability from external entities via beacons or geo-location databases. In the latter, the SU performs local sensing to determine the spectrum status, and this is commonly known as \emph{spectrum sensing}. Each type has its own limitations and challenges, and in this paper, we focus on spectrum sensing\footnote{The Federal Communications Commission (FCC) has recently obviated the need of mandatory spectrum sensing when the SU has access to geo-location databases \cite{FCC4}. However, passive awareness demands some changes to the infrastructure of the PU networks. Besides, there are many challenges with geo-location databases that need to be further explored \cite{Murty}, and spectrum sensing becomes necessary when the SU fails to access these databases. Thus, spectrum sensing still stands out as a key element in cognitive radio \cite{Saeed}.}. In addition, when the SU accesses an available band, it must periodically monitor this band to account for sudden reappearances of the PUs. This would inherently limit the throughput of the SUs, or at least degrade the quality-of-service (QoS) if it is even guaranteed.

While reliable spectrum sensing techniques are pivotal, the CRN's throughput and channel maintenance stand out as important considerations for the SUs. This has primarily motivated the introduction of multiband cognitive radio (MB-CR) paradigm, which is also referred to wideband cognitive radio. By enabling SUs to simultaneously sense and access multiple channels, this paradigm promises significant enhancements to the network's throughput. In addition, it helps provide seamless handoff from band to band, which improves the link maintenance and reduces data transmission interruptions. Nevertheless, the wideband receiver front-end architecture and the multiband spectrum sensing impose additional challenges that need to be overcome.

Another design issue in CRNs is the sensing reliability. Due to the random nature of the wireless channel, sensing may become unreliable when the SU is in a deep fade (e.g., being shadowed by a building). This triggers the hidden terminal problem where the shadowed SU may incorrectly decide to access a band because of its inability to detect the PUs even when they are active. To combat fading in wireless channels, cooperative communication provides spatial diversity gains by enabling multiple SUs to cooperate together in spectrum sensing and access \cite{Ibnkahla2, Akyildiz, Khaled1, Sahai3}. Nevertheless, it is still a challenge itself on how to share or combine the detection results among SUs.

Research on multiband cognitive radio has been scattered in the past years over many papers, and it has now reached a mature stage where a unified framework of the multiband spectrum access can be established. In this paper, we try to provide this unified framework by presenting an in-depth survey of the key aspects of multiband spectrum sensing and access. The main contributions of this paper are as follows. We define the multiband detection problem and analytically parse the recent advancements in multiband spectrum sensing techniques. In particular, we address serial sensing, parallel sensing, and wideband sensing such as wavelet sensing, compressive sensing, etc. Also, we present an overview of cooperative communications from MB-CRN perspective. In specific, we present a paradigm, namely the \emph{cooperative multiband CRNs} to compromise the advantages and the limitations of both MB-CRNs and cooperative networks. Moreover, we highlight an important relationship between spatial diversity and sampling requirements in such networks. Performance metrics are among the key factors in ensuring successful multiband spectrum access, and yet they have been scattered in the literature over many publications with no unified guide between them. This paper provides the reader with a survey of performance metrics as applied to the single-band case, then extends them to the multiband framework. Consequently, we provide an extended discussion of the different tradeoffs to be taken into account while designing some of the key parameters of MB-CRNs such as sensing time and throughput tradeoff, sensing time and detection reliability tradeoff, number of cooperating SUs and power consumption tradeoff, etc.

The paper is organized as follows. Single-band spectrum sensing techniques and cooperative CRNs are reviewed in Section II. Section III overviews the multiband spectrum sensing techniques. Section IV presents the cooperative MB-CRN paradigm. Section V provides several performance metrics for MB-CRNs, and the fundamental limits and tradeoffs in MB-CRNs are outlined in Section VI. Finally, the main conclusions are drawn in Section VII.

\section{Background}

\subsection{Single-band Spectrum Sensing}
Spectrum sensing is arguably one of the most important element in CRNs. SUs must reliably detect the presence of the PUs without causing any interference to them, and this is inherently a challenging task since the detection is done independently by SUs in order not to alter the PU's network infrastructure. The spectrum sensing problem can be described by the classical binary hypothesis testing problem as
\begin{equation}
\label{eq:SingleBandSpectrumSensingProblem}
\begin{aligned}
\mathcal{H}_0:  &~  \mathbf{y}= \mathbf{v}      \\
\mathcal{H}_1:  &~  \mathbf{y}= \mathbf{x}+\mathbf{v},
\end{aligned}
\end{equation}
where $\mathbf{y}=\big[y[1],y[2],\ldots,y[N]\big]^T$ is the received signal at the SU receiver, $\mathbf{x}=\big[x[1],x[2],\ldots,x[N]\big]^T$ is the transmitted PU signal, and $\mathbf{v}$ is a zero-mean additive white Gaussian noise (AWGN) with variance $\sigma^2\mathbf{I}$ (i.e. $\mathbf{v}\sim\mathcal{N}(\mathbf{0},\sigma^2\mathbf{I})$), where $\mathbf{I}$ is the identity matrix. Note that $\mathcal{H}_0$ and $\mathcal{H}_1$ indicate the absence and the presence of the PU, respectively. Typically, to decide between the two hypotheses, we compare a test statistic with a predefined threshold, $\lambda$. Mathematically, this is written as
\begin{equation}
\label{eq:LRT}
T(\mathbf{y})    \overset{\mathcal{H}_1}{\underset{\mathcal{H}_0}{\gtrless}} \lambda,
\end{equation}
where $T(\mathbf{y})$ is the test statistic (usually the likelihood ratio test (LRT) \cite[ch.~II]{Poor}). There are two design elements in spectrum sensing. First, formulating a proper test statistic that reliably gives correct information about the spectrum occupancy. Second, setting a threshold value that differentiates between the two hypotheses. The most common single-band sensing techniques are: coherent detection, energy detection, and feature detection.

\subsubsection{Coherent Detection}
When the SU has a perfect knowledge of the PU signal structure, it can correlate the received signal with a known copy of the PU signal. The test statistic is given by
\begin{equation}
\label{eq:CoherentDecisionMetric}
T(\mathbf{y})    =   \Re\big[\mathbf{x}^H\mathbf{y}\big],
\end{equation}
where $\Re$ is the real part, and $(.)^H$ is the Hermitian (conjugate-transpose) operator.

\subsubsection{Energy Detection}
When the SU does not have prior knowledge of the PU's transmitted signal, the energy detector can be used. It simply computes the energy of the received signal over a time window. The test statistic for a typical energy detector is expressed as
\begin{equation}
\label{eq:EnergyDecisionMetric}
T(\mathbf{y})    =   \frac{1}{\sigma^2} ||\mathbf{y}||^2,
\end{equation}
where $||.||$ is the Frobenius norm.

\subsubsection{Feature Detection}
In practical wireless communication systems, the transmitted signals are deliberately embodied by some unique features to assist the receiver in detection \cite{Ibnkahla1}. These features are due to the redundancy added to the transmitted signal, and they provide better robustness against noise uncertainties \cite{Ghaith2}. These features could be detected from the second-order statistics. For example, it is sufficient, under certain conditions, to compute the second-order statistics of the PU transmitted signal. That is, the test statistics is
\begin{equation}
\label{eq:2ndMomentMetric}
T(\mathbf{y})   =   E\big[\mathbf{y}\mathbf{y}^H\big].
\end{equation}
If the PU signal has periodic statistical properties, then instead of using the \emph{power spectral density} (PSD), we can use the \emph{cyclic spectral density} (CSD). For instance, the CSD of the received signal is expressed as \cite{Gardner1}
\begin{equation}
\label{eq:CSD}
S_y(f,\alpha)=\sum_{\tau=-\infty}^{\infty}R_{y}^{\alpha}(\tau)e^{-j2\pi f \tau},
\end{equation}
where $R_{y}^{\alpha}(\tau)$ is the Fourier coefficients of the autocorrelation function, and $\alpha$ is the cyclic frequency.
The binary hypothesis testing problem in (\ref{eq:SingleBandSpectrumSensingProblem}) becomes
\begin{equation}
\label{eq:CyclicDetectionProblem}
\begin{aligned}
\mathcal{H}_0:  &~  S_y(f,\alpha) = S_v(f,\alpha)                      \\
\mathcal{H}_1:  &~  S_y(f,\alpha)  = S_x(f,\alpha)+S_v(f,\alpha).
\end{aligned}
\end{equation}
Detectors that solve this problem are known as the \emph{cyclostationarity detectors}.
Other detectors can exploit the eigenvalue structure of the covariance matrix of PU signals. For instance, in \cite{Zeng1}, it is demonstrated that the covariance matrix of the transmitted signal and noise are not alike. Similarly, in \cite{Zeng3}, the test statistic is simply the ratio of the maximum and minimum eigenvalues of the PU signal's sample covariance matrix where prior knowledge of the PU signal is not required.

An overall comparison between these techniques is provided in Table \ref{tab:SingleBandDetectors}. Each detector has its advantages and limitations. The choice of the detector depends on many factors such as how much information the SU has \emph{a priori} about the PU signal. For example, coherent detection is preferred when the SU has full knowledge (e.g. bandwidth, carrier frequency, modulation, packet format, etc.). It quickly achieves high processing gain (i.e. it requires fewer samples compared to other detectors) \cite{Ghaith2}. However, since the SU must detect different bands in the spectrum, it requires knowledge of each signal structure of these bands which is usually infeasible for the SU to obtain. Feature detectors could be used when partial knowledge is available (e.g. pilots, cyclic-prefixes, preambles, etc.). Such detectors are robust against noise uncertainties. Nevertheless, they demand more processing complexity, sensing time, and power consumption compared to other detectors. Finally, the energy detector is very simple and it does not require any prior knowledge of the PU signal. However, the threshold depends on the noise variance, $\sigma^2$, and if it is not accurately estimated, the performance of this detector becomes very poor. In fact, uncertainties in $\sigma^2$ make detection impossible below a signal-to-noise ratio (SNR) level known as the \emph{SNR wall} \cite{Sahai1}. Here, the SNR denotes the signal power at the SU side (i.e. the power of $y_i$) to the noise power ratio.

\begin{table*}[!tp]
\renewcommand{\arraystretch}{1.3}
\small
\caption{A comparison between the common single-band spectrum sensing techniques}
\label{tab:SingleBandDetectors}
\centering
\begin{tabularx}{\textwidth}{lcccccc}
\toprule
\multirow{2}[0]{*}{\bfseries{Sensing Technique}} & \multicolumn{2}{c}{\bfseries{Required Information}} & \multicolumn{2}{c}{\bfseries{Distinguish PU from}} & \multirow{2}[0]{*}{\bfseries{Limitations}} \\
& $\sigma^2$     & $\mathbf{x}$      & \bfseries{Noise}          & \bfseries{Other Signals}     &                                \\\toprule
\bfseries{Coherent Detector}          & No      & Yes          & Yes            & Yes               &Requires synchronization    \\
\bfseries{Energy Detector}            & Yes     & No           & SNR dependent  & No                &Performance depends on noise power estimation\\
\bfseries{Cyclostationarity}         & No      & Yes          & Yes            & Yes               &Complex processing and high sensing time         \\
\bfseries{$2^{nd}$ Moment Detector}   & No      & Yes          & Yes            & Yes               &Applicable merely for Gaussian signals                          \\
\bfseries{Covariance Detector}        & No      & No           & Yes            & Yes               &Some test statistics require knowledge of $\mathbf{x}$\\
\bottomrule
\end{tabularx}%
\end{table*}

\subsection{Cooperative Cognitive Radio Networks}
\begin{figure}[!b]
\centering
\includegraphics[width=3.5in]{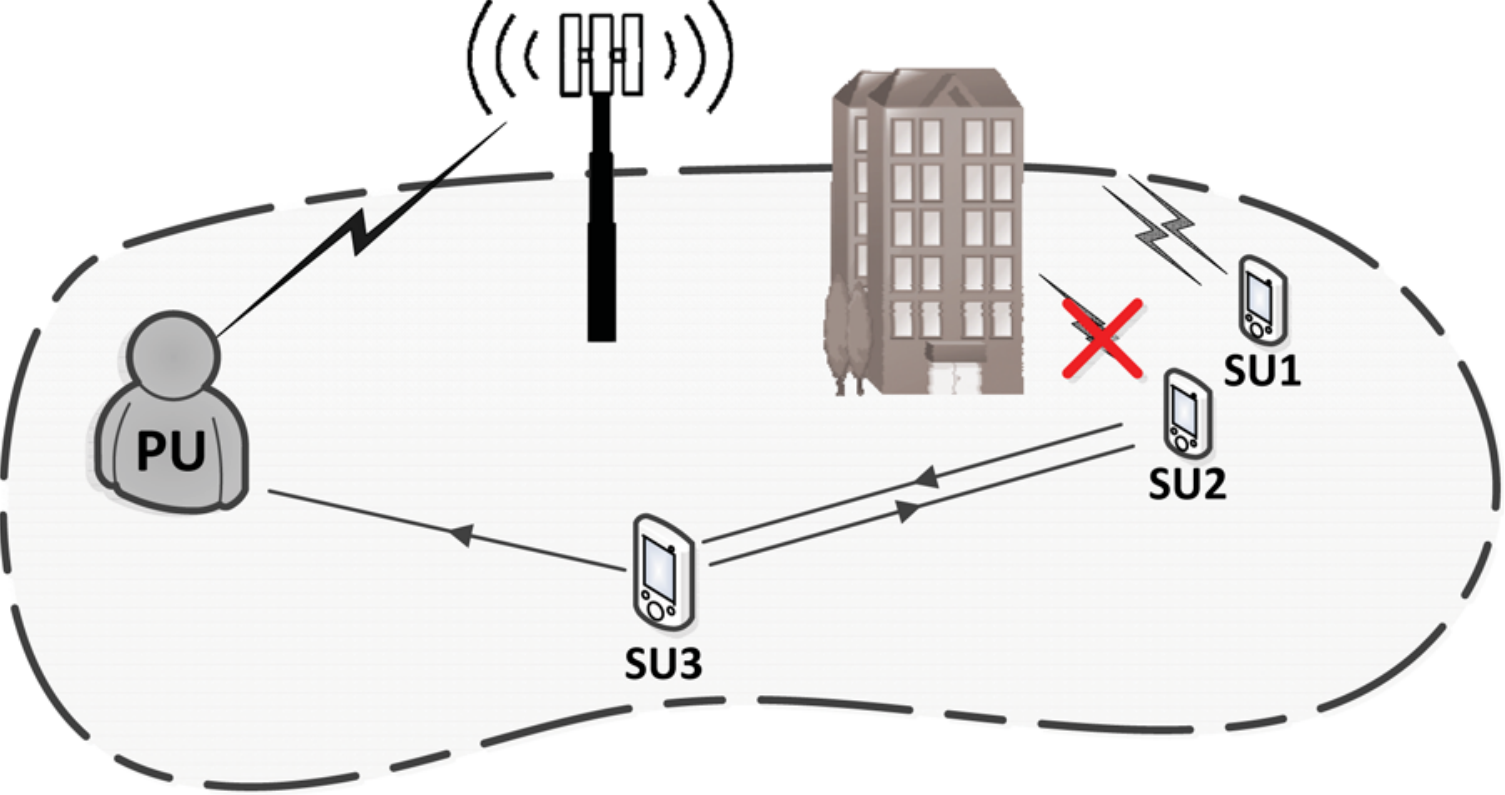}
\caption{Cooperation between SUs helps mitigate the hidden terminal problem.}
\label{fig:Hatta1}
\end{figure}

One of the common challenges in wireless communication systems is the hidden terminal problem. This phenomenon arises due to the random nature of the wireless channel where the SU could be shadowed by an object or in a deep fade. For instance, in Fig. \ref{fig:Hatta1}, SU1 is shadowed by a building, and hence it would decide that there is a spectrum opportunity even though the PU is present. In contrary, SU2 collaborates with SU3, and it realizes the presence of the PU via the information shared from SU3, and hence it does not access the spectrum to avoid interference with the PU. This is the basic principle of cooperative sensing where the SUs in a certain geographical region would cooperate together to effectively improve the sensing reliability \cite{Ibnkahla2, Akyildiz, Khaled1, Sahai3}.

One of the key issues in cooperative communications is how to combine the collected information from the participating SUs. There are three main techniques, namely: hard combining, soft combining, and hybrid combining.

\subsubsection{Hard Combining}
In this technique, the SU merely sends its final one-bit decision to the other SUs. If we have $K$ cooperating SUs, the final decision metric is expressed as \cite{Wei}
\begin{equation}
\label{eq:HardMetric}
\mathcal{D}=\sum_{i=1}^{K}d_i\left\{
\begin{array}{lr}
<k \text{, }            &  \mathcal{H}_0    \\
\geq k \text{, }        & \mathcal{H}_1
\end{array} \right.,
\end{equation}
where $d_i \in$ $\{0,1\}$ is the final decision made by the $i$-th SU such that `0' and `1' indicate the absence and the presence of the PU, respectively. This is basically a logical decision metric such that:
\begin{itemize}
\item If $k=1$, then (\ref{eq:HardMetric}) is an OR-logic rule (i.e. the PU is considered present if only \emph{one} SU sends `1').
\item If $k=K$, then (\ref{eq:HardMetric}) is an AND-logic rule (i.e. the PU is considered present if \emph{all} the SUs send `1').
\item If $k=\lceil K/2\rceil$ ($\lceil x\rceil$ denotes the smallest integer not less than $x$), then (\ref{eq:HardMetric}) becomes a majority rule (i.e. the PU is considered present if the majority of the SUs send `1').
\end{itemize}
Note that the OR-logic rule guarantees minimum interference to the PU since only a single `1' is enough to declare the band occupied, whereas the AND-logic rule guarantees higher throughput since the band is considered occupied by the PU when all SUs send `1'. For example, referring back to Fig. \ref{fig:Hatta1}, we have $K=2$ where SU2 sends a `0' and SU3 sends a `1'. Thus, if AND-logic is used, the PU would be declared absent, and if OR-logic is used, then the PU would be declared present because the `1' sent by SU3 is enough.

\subsubsection{Soft Combining}
In this technique, the SU shares its original sensing information (or original statistics) with the other SUs without locally processing them. It is shown that the optimal combination (OC) is actually based on the weighted summation of the observed statistics from the collaborative SUs \cite{Jun1}. Mathematically, this is written as
\begin{equation}
\label{eq:SoftCombining}
\mathcal{T}   =   \sum_{k=1}^{K}  c_k T_k(\mathbf{y}),
\end{equation}
where $T_k(\mathbf{y})$ is the $k$-th user test statistics, and $c_k$ is the weight coefficient. These weights could be based on equal gain combination (EGC) or maximal ratio combination (MRC). In the former, $c_k=1$, and in the latter, $c_k$ is proportional to the SNR of the link between the PU and the $k$-th user. It is demonstrated that the OC converges to EGC scheme in high SNR, whereas it converges to MRC in low SNR region \cite{Jun1}.

\subsubsection{Hybrid Combining}
Hard combining requires less overhead compared to soft combining. However, since the statistics at each SU is reduced to one-bit, there is an information loss that propagates to the other SUs. Therefore, the final decision is less reliable compared to soft combing. This has motivated the authors in \cite{Jun1} to propose a \emph{softened hard} combining scheme where the SU sends two-bits overhead instead of one-bit. This provides a good balance between the hard and soft combination schemes.
In general, increasing the number of bits will improve the performance at the expense of larger overhead. Further studies are required to determine the optimal number of bits to meet a certain detection performance.

\section{Multiband Spectrum Sensing}
In this section, we describe the multiband detection problem and its typical applications. Then, we analyze the recent advancements of multiband spectrum sensing techniques.

\subsection{The Multiband Detection Problem}
Multiband cognitive radio networks (MB-CRNs) have recently caught the attention since they can significantly enhance the SUs' throughput. There are several scenarios where MB-CRN can be encountered such as:
\begin{itemize}
\item Many modern communication systems and applications require a wideband access. The wideband spectrum can be divided into multiple subbands or subchannels. Thus, the problem becomes a multiband detection problem.
\item When a SU wants to minimize the data interruptions due to the return of PUs to their bands, seamless handoff from one band to another becomes vital. Therefore, the SU must have backup channels besides those channels it is currently using. With MB-CRNs, the SU does not only have a set of candidate channels, but it can also reduce handoff frequency.
\item When a SU wants to achieve higher throughput or maintain a certain QoS, then it may transmit over a larger bandwidth, and this is primarily enabled by accessing multiple bands.
\item In cooperative communications, multiple SUs may share their detection results among each other. However, if each SU monitors a subset of subchannels, and then shares its results with others, then the entire spectrum can be sensed, and consequently, more opportunities are explored for spectrum access.
\end{itemize}

\begin{figure}[!b]
\centering
\includegraphics[width=3.5in]{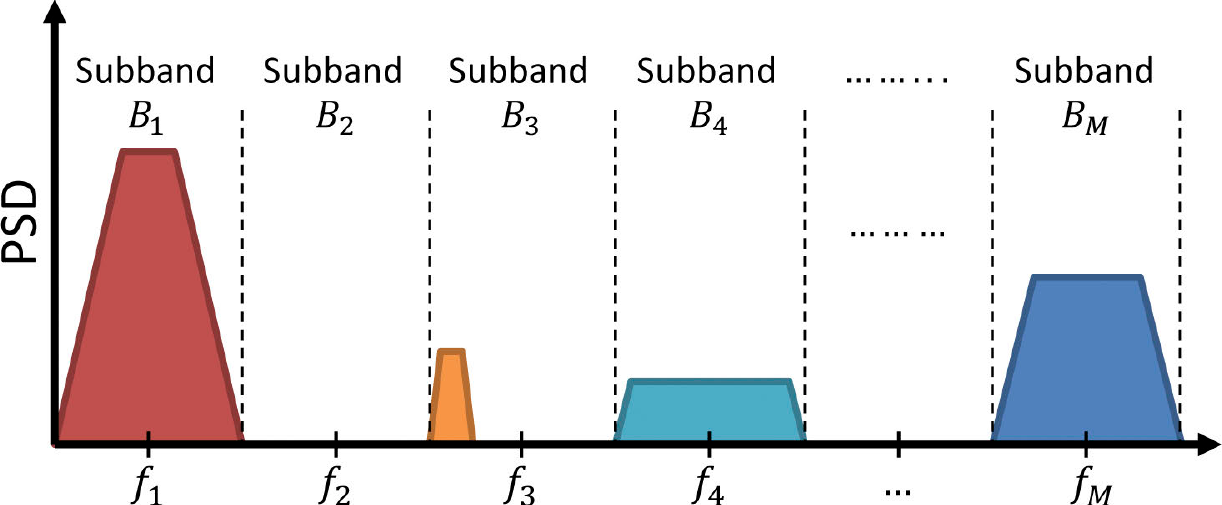}
\caption{A wideband spectrum can be divided into non-overlapping subbands.}
\label{fig:Hatta2}
\end{figure}

Consider Fig. \ref{fig:Hatta2} where we assume that we can divide the wideband spectrum into $M$ non-overlapping subchannels (or subbands). For simplicity, we assume that each subband has the same bandwidth. Clearly, the SU's primary task is to determine which subchannels are available for spectrum access. This is, in general, a challenging task since the available bands are not necessary contiguous, and the activity of the PUs might be correlated across these bands (e.g., the PUs in WLAN and broadcast television \cite{Hossain}). In addition, each particular band is considered occupied even if a small portion of it is only being used. For example, in IEEE 802.22, the 6MHz channel must not be accessed, under the interweave paradigm, when it is being used by a wireless microphone that consumes only 200KHz \cite{GwangZeen}. In Fig. \ref{fig:Hatta2}, a PU occupies a small portion of subband $B_3$, and hence it must not be used by the SU when it follows the interweave paradigm.

If we assume that the subbands are independent, then the MB sensing problem reduces to a binary hypothesis for each one. Mathematically, this is expressed as
\begin{equation}
\label{eq:MultibandDetectionProblem}
\begin{aligned}
\mathcal{H}_{0,m}:  &~  \mathbf{y}_m= \mathbf{v}_m               &,m=1,2,\ldots,M      \\
\mathcal{H}_{1,m}:  &~  \mathbf{y}_m= \mathbf{x}_m+\mathbf{v}_m, &,m=1,2,\ldots,M
\end{aligned}
\end{equation}
where individual subband $m$ is indicated by subscript $m$. The decision rule for each band is
\begin{equation}
\label{eq:MBDecisionRule}
T(\mathbf{y}_m)    \overset{\mathcal{H}_{1,m}}{\underset{\mathcal{H}_{0,m}}{\gtrless}} \lambda_m.
\end{equation}
While SB sensing constitutes the building block of MB spectrum sensing, many modifications and advancements are required to put SB sensing into feasible implementation for MB sensing. In the following section, we present three main sensing techniques for MB access.

\subsection{Serial Spectrum Sensing Techniques}
In serial spectrum sensing, any of the aforementioned SB detectors can be used to sense multiple bands one at a time using any of the following techniques:

\subsubsection{Reconfigurable Bandpass Filter (BPF)}
A reconfigurable bandpass filter (BPF) can be implemented at the receiver front-end to pass one band at a time, and then a SB detector is used to determine whether that particular band is occupied or not, as shown in Fig. \ref{fig:Hatta3a}. Clearly, this requires a wideband receiver front-end, which sets several challenges for hardware implementation due to the high sampling rates. In addition, controlling the cutoff frequency and the filter's bandwidth are challenging design issues \cite{Joshi}.

\subsubsection{Tunable Oscillator}
Another approach is based on tunable local oscillator (LO) that down-converts the center frequency of a band to a fixed intermediate frequency as shown in Fig. \ref{fig:Hatta3b}. This will significantly reduce the sampling rate requirement.

The main limitation of reconfigurable BPFs and tunable oscillators is that they require tuning and sweeping as the sensing moves from one channel to the next. This hinders fast processing, and thus these techniques are undesirable.

\subsubsection{Two-Stage Sensing}
\begin{figure}[!b]
\centerline{\subfigure[]{\includegraphics[width=3.5in]{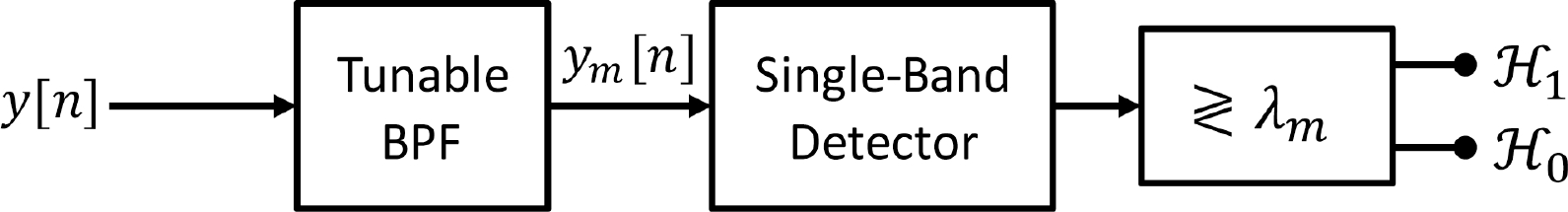}
\label{fig:Hatta3a}}}
\centerline{\subfigure[]{\includegraphics[width=3.5in]{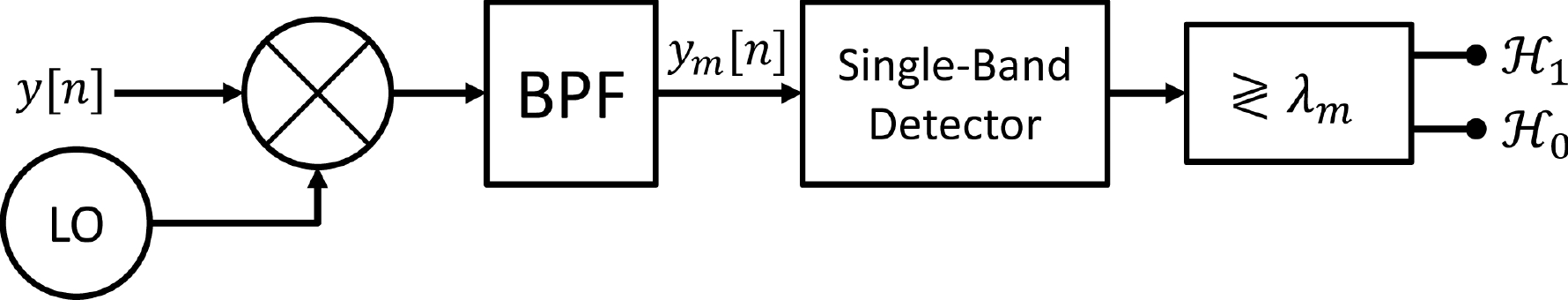}
\label{fig:Hatta3b}}}
\centerline{\subfigure[]{\includegraphics[width=3.5in]{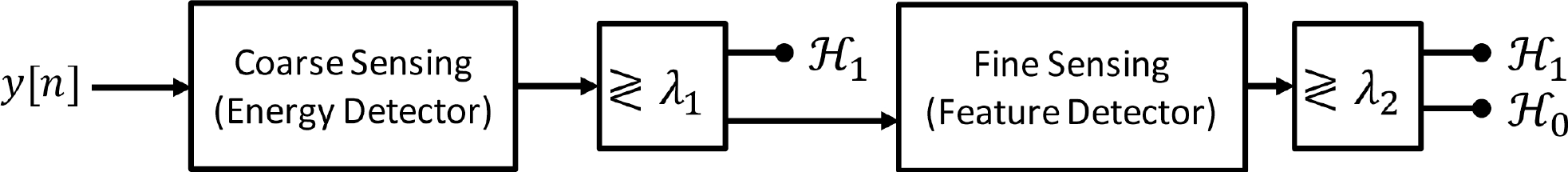}
\label{fig:Hatta3c}}}
\caption{(a) A tunable BPF; (b) A local oscillator; (c) A two-stage serial spectrum sensing scheme.}
\label{fig:Hatta3}
\end{figure}

Spectrum sensing can be done over two stages. A coarse sensing is first performed followed by a fine sensing stage if necessary \cite{Luo,Maleki}. The block diagram of this scheme is presented in Fig. \ref{fig:Hatta3c}. For instance, in \cite{Luo}, an energy detector is used in both stages. In the coarse stage, a quick search is done over a wide bandwidth, and in the fine stage, the sensing is done over the individual candidate subbands in that bandwidth, one at a time. Simulation results show that two-stage spectrum sensing provides faster searching time compared to one-stage based searching algorithms when the PU activity is high. In \cite{Maleki}, the coarse stage is based on energy detection due to its fast processing. If the test statistics is larger than a predefined threshold, then the band is considered occupied. Otherwise, a fine stage is performed where a cyclostationarity detector is implemented due to its robustness at the low SNR regime.

\subsubsection{Other Algorithms}
There are several other techniques that are used to serially sense multiple bands such as sequential probability ratio tests (SPRTs). SPRTs have been extensively used to provide efficient, yet fast channel search algorithms. Unlike conventional test statistics that use a fixed number of samples, SPRTs tend to reduce the average number of samples required to be collected to achieve a certain performance. The basic principle is to collect samples as long as $a<T(\mathbf{y})<b$, where $a$ and $b$ are predetermined bounds. A decision will be made once the test statistic is outside these bounds (particularly, we decided $\mathcal{H}_0$ if $T(\mathbf{y})<a$ and $\mathcal{H}_1$ if $T(\mathbf{y})>b$)  \cite[ch.~III]{Poor}. For instance, Dragalin proposed in \cite{Dragalin} a SPRT searching algorithm where it is assumed that there is only one available channel for spectrum access. This assumption implies that the channel occupancy is correlated as well as the SU spectrum access is eventually limited to one channel. In \cite{Lai}, a Bayesian-based SPRT is adopted, yet it is usually infeasible to implement since the Bayesian framework requires both prior probabilities of the PU signals as well as some cost structures \cite[ch.~II]{Poor}. Also, it is not possible to retest the channels again because the authors assume that the number of channels are infinite, which is not practical for CRNs. The limitations of these works have motivated the work in \cite{YanXin} where it is assumed that the number of channels is finite and no specific cost structures are required. Two efficient algorithms are analyzed based on SPRT and energy detection, and both algorithms reduce the sensing time compared to Dragalin's algorithm. Moreover, an agile multi-detector, called \emph{iDetector}, is proposed in \cite{Ejaz}. This detector intelligently sets a detection algorithm based on the availability of PU signal information at the SU side. For example, if no information can be obtained for some bands, then a combination of energy and cyclostationarity detection is used, and for bands where such information are easier to obtain, the coherent detector can be used. The detection reliability is comparable with the cyclostationarity detector, yet the detection time is significantly lower. Obviously, integrating all these detectors in one receiver imposes higher costs and complexity.

In general, serial spectrum sensing has high average searching time when the presence probability of the PU is high \cite{YanXin}. This has motivated the researchers to come up with more advanced receivers for MB spectrum sensing.

\subsection{Parallel Spectrum Sensing}
\begin{figure}[!tb]
\centerline{\subfigure[]{\includegraphics[width=3.5in]{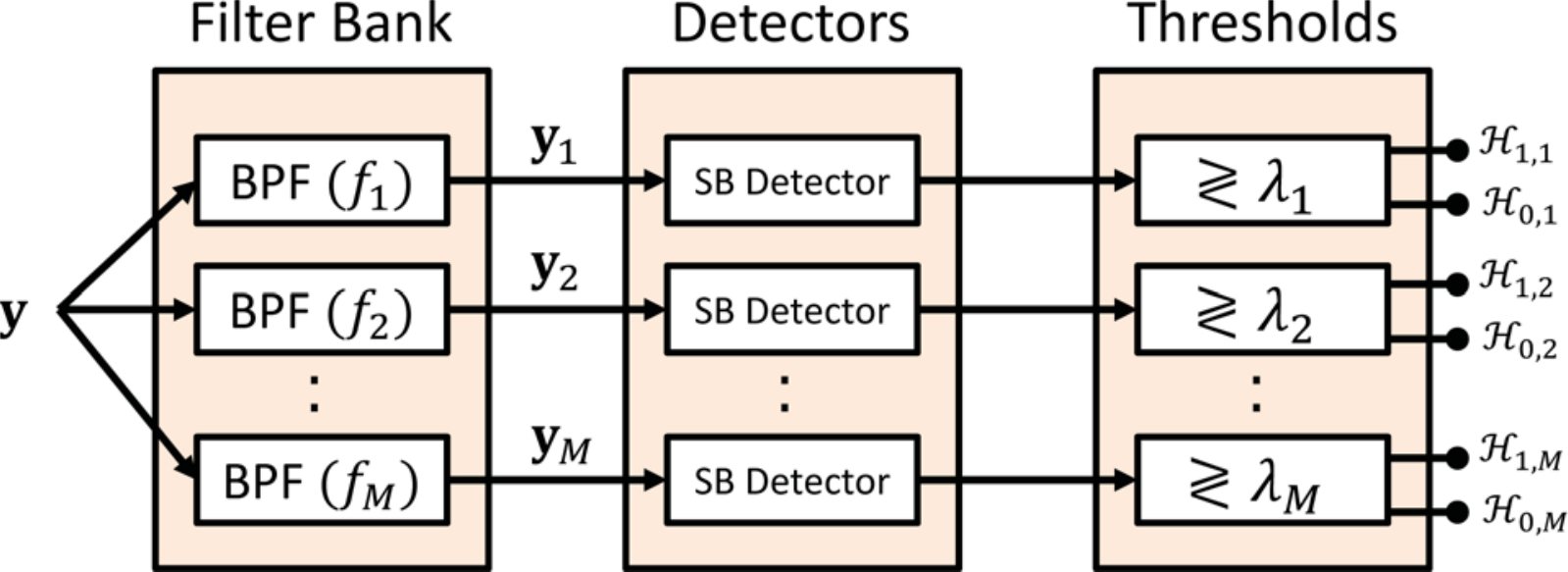}
\label{fig:Hatta4a}}}
\centerline{\subfigure[]{\includegraphics[width=3.5in]{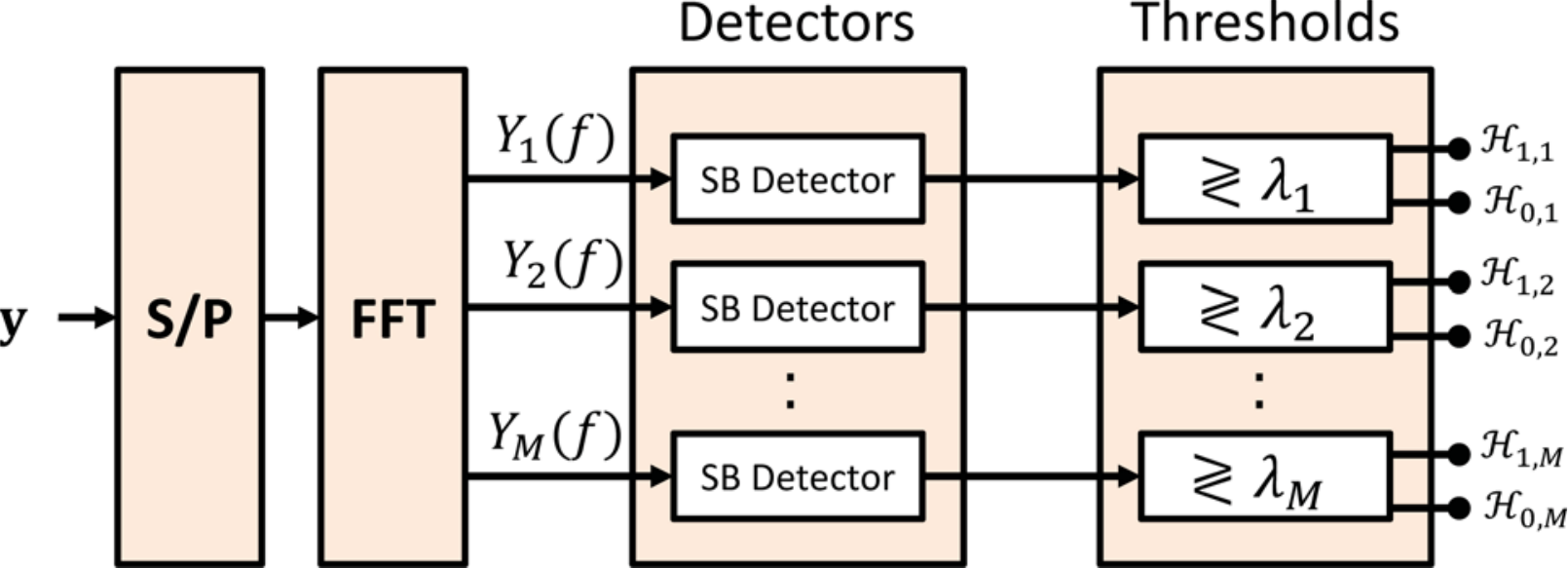}
\label{fig:Hatta4b}}}
\caption{(a) A filter bank structure; (b) Frequency-based parallel SB detectors.}
\label{Hatta4}
\end{figure}

In parallel sensing, the SU is equipped with multiple SB detectors such that each one senses a particular band. This can be done using a filter bank as shown in Fig. \ref{fig:Hatta4a}. It consists of multiple BPFs each with a certain center frequency followed by SB detectors \cite{Farhang}. Even though this filter bank merely considers one-type of multiple SB detectors (i.e. \emph{homogenous} structure), we can extend this principle to have \emph{heterogenous} structure with different multiple SB detectors. For instance, since pilots are being used in TV bands, we can use multiple SB feature detectors over these bands. For bands with unknown signal structures, energy detectors can be implemented. However, filter banks demand many RF components which make implementation expensive and the receive size larger. The complexity will also be further increased if the detectors are not of the same type.

The wideband spectrum can be decomposed in the frequency domain as shown in Fig. \ref{fig:Hatta4b}. This is done using a serial-to-parallel (S/P) converter and Fast Fourier Transform (FFT) before feeding the signal to SB detectors. The energy detector is the most commonly used technique here because it is easy to compute the energy in the frequency domain by analyzing the power spectral density (PSD) \cite{Quan1, Beaulieu1, Farooq, Beaulieu2, Shi, Hossain}. This can be expressed as
\begin{equation}
\label{eq:PSD}
T(\mathbf{y}_m) = \sum_{n=1}^{N} |Y_m(n)|^2, ~m=1,2,\ldots,M,
\end{equation}
where $Y_m(n)$ is the frequency domain representation of the received signal $\mathbf{y}$ at the $m$-th subchannel, and $N$ here is interpreted as the FFT size.
Clearly, each subchannel has its own threshold (in a vector form we have $\boldsymbol {\lambda}=[\lambda_1,\lambda_2,\ldots,\lambda_M]$). In \cite{Quan1}, Quan et al. propose a \emph{multiband joint detector} (MJD) to maximize the CRN's throughput. It is demonstrated that when the thresholds are jointly optimized, significant throughput gains can be attained compared to a uniform threshold approach (i.e. $\boldsymbol {\lambda}=\lambda\mathbf{1}$). The proposed algorithm intelligently assigns higher thresholds for the bands with higher opportunistic rate, and lower thresholds for the bands that require higher PU protection. The former helps minimize transmission interruptions, and the latter helps reduce interference with PUs.

The MJD has become a benchmark in wideband spectrum sensing, and many works have been recently proposed to improve it. In particular, the MJD lacks the periodic sensing, an important requirement in spectrum sensing, and hence the authors in \cite{Beaulieu1} have proposed a \emph{multiband sensing-time-adaptive joint detector} (MSJD) where a dynamic sensing time is proposed. It is shown that using dynamic sensing time can remarkably improve the throughput of the network. In \cite{Beaulieu2}, the authors present a cost and time effective MJD to reduce the system's complexity. In addition, a modified MJD is proposed in \cite{Iqbal} where the energy detectors are replaced by coherent detectors.

The previous techniques assume that the subchannels are independent, which is not generally the case in practice since the subchannels may be correlated. This may arise when the PU transmits over multiple channels, so the occupancy of one channel is correlated with the neighboring channels. Another scenario arises when the PU uses high power in one of the channels such that the neighboring channels experience adjacent channel interference (ACI), and this leads to some correlation among this set of channels. In \cite{Axell3}, the authors investigate the impact of noise power uncertainties when it is correlated between the subchannels. It is demonstrated that losing the independency between the subchannels makes the complexity of the detection problem to exponentially increase as the number of subchannels increases. To mitigate such high complexity, the authors in \cite{Hossain} propose a linear energy combiner where (\ref{eq:PSD}) becomes
\begin{equation}
\label{eq:CorrelatedPSD}
T(\mathbf{y}_m) = \sum_{n=1}^{N} w_n |Y_m(n)|^2, ~m=1,2,\ldots,M,
\end{equation}
where $\{w_n\}$ are weighting coefficients that must be optimized. It is demonstrated that this energy combiner outperforms the MJD in terms of detection reliability when the occupancy of the bands are correlated. Nevertheless, the authors assume that the correlation model is known \emph{a priori}. Therefore, for practical implementations, further work is required to come up with efficient algorithms that are robust against different subchannel correlation models.

In the following, we discuss recent advancements in multiband detectors, and particularly we will investigate the wavelet sensing (WS), compressive sensing (CS), angle-based sensing (AS), and blind sensing (BS).

\subsection{Wavelet Sensing}
One of the assumptions that have been made in the aforementioned techniques is that the SU knows the number of subbands, $M$, and their corresponding locations at $f_1, f_2, \ldots, f_M$. However, in practice, this assumption is not very practical since the CRN must be able to support heterogeneous technologies that have different requirements (e.g., transmission schemes, bandwidth, etc.). To overcome this problem, wavelet based-detectors have become a good candidate due to their ability to detect and analyze the singularities in the spectrum \cite{Mallat}. These singularities have important interpretations since they occur at the edges of the subbands (i.e. when we transit from one band to the neighboring bands). Tian and Giannakis have used the wavelet transform for MB spectrum sensing in \cite{Tian1}, where they propose a continues wavelet transform (CWT) that is carried out in the frequency domain to detect the singularities of the wideband spectrum. In other words, using the CWT, the authors have successfully determined the boundaries of the subbands without prior knowledge of the number of subbands and their corresponding center frequencies. Once these edges are determined, the PSD is estimated to determine which subchannels are vacant for opportunistic access. This type of spectrum sensing is referred as \emph{edge detection}.

\begin{figure}[!t]
\centering
\includegraphics[width=3.5in]{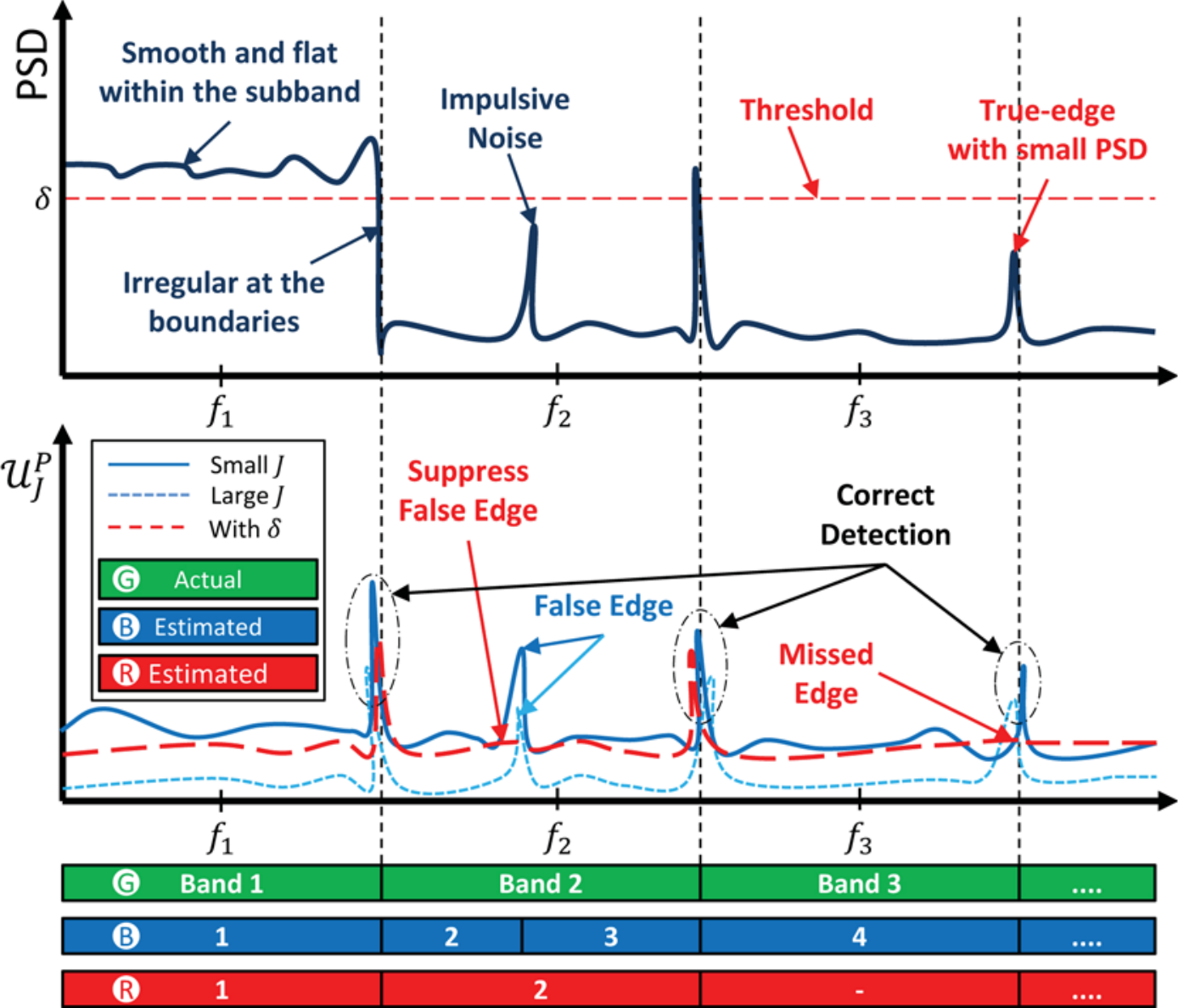}
\caption{Increasing $J$ enhances the detection and suppresses noise, yet impulsive noise degrades the accuracy of boundary estimation. Thus, using a threshold helps ignore such false edges, yet it may miss actual boundaries.}
\label{fig:Hatta5}
\end{figure}

Mathematically, the CWT is expressed as
\begin{equation}
\label{eq:WaveletTransform}
\mathcal{W}_s(f)   =   S(f) * \psi_s(f),
\end{equation}
where $S(f)$ is the wideband PSD as a function of frequency, $*$ is the convolution operator, and
\begin{equation}
\label{eq:WaveletSmoothingFunction}
\psi_s(f)   =   \frac{1}{s} \psi\bigg(\frac{f}{s}\bigg),
\end{equation}
where $\psi(f)$ is called the \emph{wavelet smoothing function}, and $s$ is called the \emph{dilation factor}. The authors in \cite{Tian1} have used the derivatives of the CWT since they sharpen the edges to help better characterize them. This method is known as the \emph{wavelet modulus maxima} (WMM). To further enhance the peaks caused by the edges and to suppress noise, the \emph{wavelet multiscale product} (WMP) can be used. It is simply the product of the $J$ first-derivative CWTs, and it is expressed as
\begin{equation}
\label{eq:MultiscaleProductWavelet}
\mathcal{U}_J^M   = \prod_{j=1}^{J} \mathcal{W}'_{s(j)}(f),
\end{equation}
where $\mathcal{W}'_s(f)$ is the first derivative of $\mathcal{W}_s(f)$ in (\ref{eq:WaveletTransform}), and $s(j)=2^j$. It must be noted that increasing $J$ further improves the reliability of edge detection, and this is at the expense of additional complexity. One of the challenges of this technique, however, is that these sharp edges do not merely arise at the boundaries of the subbands but also arise due to other sources (e.g., impulsive noise and spectral leakage). These undesired edges may degrade the boundary estimation. For example, consider the spectrum shown in Fig. \ref{fig:Hatta5}. There are three edges at the boundaries of $f_1$, $f_2$, and $f_3$ bands, and one another edge due to the impulsive noise. The three true-edges are correctly detected using WMP, and if a larger number of products is taken (i.e. larger $J$), then the estimation is further improved. However, the edge estimation is not completely correct because the impulsive noise provides a false-edge, and thus the edge estimation is incorrect (i.e. the blue bar shows 4 bands instead of 3 bands). To alleviate false-edges, Zeng et al. propose a robust algorithm in \cite{Zeng2} where the local maxima of (\ref{eq:MultiscaleProductWavelet}) are compared to a threshold, $\delta$, to limit the number of edges. That is, a local maximum will only be counted as an edge if it is larger than $\delta$. Otherwise, it will be neglected. Since the local maximum depends on the shape of the wavelet and the PSD at that point, then $\delta$ is not fixed. To mitigate this variability, the WMP in (\ref{eq:MultiscaleProductWavelet}) is normalized by the mean of the PSD. However, since a threshold is introduced, the SU may miss an actual subchannel boundary when this boundary is heavily corrupted by noise (i.e. the local maximum at this boundary would be less than $\delta$). So referring back to Fig. \ref{fig:Hatta5}, when the threshold is used, the false-edge can be ignored, but if an actual edge has a PSD below $\delta$ such as the edge at $f_3$ band, then it will be missed too, and thus the estimation will be degraded (see the red bar).

Another algorithm is proposed in \cite{Xu} where the \emph{wavelet multiscale sum} (WMS) is used instead of WMP. That is, (\ref{eq:MultiscaleProductWavelet}) becomes
\begin{equation}
\label{eq:MultiscaleSumWavelet}
\mathcal{U}_J^S   = \sum_{j=1}^{J} \mathcal{W}'_{s(j)}(f).
\end{equation}
The reason of using the summation over the product is that narrowband signals with slow variations of the PSD are not detected by the multiscale product because they are attenuated when the multiplication operation is used. In such conditions, the multiscale sum is shown to provide a better performance \cite{Xu}. While it is believed that increasing $J$ in WMS smooths the edges \cite{Xu}, the authors in \cite{ElKhamy} demonstrate that the reason of this degradation is attributed to the orthogonal wavelet family used in \cite{Xu}, and to alleviate this degradation (or smoothing), non-orthogonal smoothing functions must be implemented instead. This is at the expense of higher miss detection in low SNR regions. Finally, in \cite{Hur}, a two-stage sensing is proposed: A coarse sensing, where the wavelet transform is implemented to identify the set of candidate subchannels followed by a fine-sensing stage that exploits signal features to determine which subchannels are unoccupied.

To summarize, WMM, WMP, and WMS each have their own advantages and disadvantages. Further advancements are required to provide a robust algorithm that successfully detects subband edges and neglects false edges at low complexity. Also, different smoothing functions (orthogonal and non-orthogonal) must be studied to analyze their impact on the quality of edge detection.

\subsection{Compressive Sensing}
Conventionally, to successfully reconstruct the received signal, the sampling rate must be at least as twice the maximum frequency component in the signal (also known as the Nyquist rate) \cite{Shannon}. For instance, if the wideband spectrum of interest has a 3GHz bandwidth, then the sampling rate must be at least 6GHz, which is very challenging in terms of feasible implementation and signal processing. Thus, one may ask: can we sample below this rate, and yet successfully recover the signal? The answer is yes, under certain conditions, and this is enabled by compressive sampling (CS) (also known as compressive sensing).

Compressive sensing has become an active area of research due to its capability to tangibly reduce the sampling rate when the signal is \emph{sparse} in a certain domain \cite{Donoho, Baraniuk, Candes}. For example, signal sparsity in frequency domain indicates that the signal has relatively less significant frequency components compared to its bandwidth. In other words, it has a lower information rate (how much information the signal has) compared to its Nyquist rate, and it is shown in \cite{Kirolos1} that the number of samples is proportional to the signal information rate not to its Nyquist rate. Since the wideband spectrum is underutilized, or in other words, it is sparse in frequency domain (recall that this is the main motivation for introducing CR), then CS appears to be a good candidate for MB spectrum sensing \cite{Tian2, Tian3, Tian4, Tian5, Polo1}.

Mathematically, assume that the received $N\times1$ discrete-time signal $\mathbf{y}$ can be written as
\begin{equation}
\label{eq:CompressiveSensing1}
\mathbf{y}  =   \mathbf{\Psi} \mathbf{s},
\end{equation}
where $\mathbf{\Psi}$ is an $N \times N$ \emph{sparsity basis matrix}, and $\mathbf{s}$ is an $N\times1$ weighting vector. The signal $\mathbf{y}$ is said to be $L$-sparse if it can be represented by linear combination of only $L$ basis vectors (i.e. only $L$ elements in $\mathbf{s}$ are nonzero) \cite{Baraniuk}. For $L\ll N$, $\mathbf{y}$ is said to be \emph{compressible} if it has few large coefficients and the rest are small or zero coefficients. The compressive sensing problem can be described by \cite{Baraniuk}
\begin{equation}
\label{eq:CompressiveSensing2}
\mathbf{z}  =   \mathbf{\Phi} \mathbf{y}    =  \mathbf{\Phi} \mathbf{\Psi} \mathbf{s},
\end{equation}
where $\mathbf{z}$ is an $O\times1$ \emph{measurement vector}, $\mathbf{\Phi}$ is an $O \times N$ \emph{measurement matrix} and is non-adaptive (i.e. its columns are fixed and independent of $\mathbf{y}$). The CS problem is to design a stable $\mathbf{\Phi}$ such that when we reduce the dimension of $\mathbf{y}\in \mathbb{R}^N$ to $\mathbf{z}\in \mathbb{R}^O$, we do not lose the signal information. Also, we need a reconstruction algorithm to recover $\mathbf{y}$ from only $O\approx L$ measurements of $\mathbf{z}$. Note that since $O<<N$, we have infinitely many solutions to (\ref{eq:CompressiveSensing2}), and hence there are several existing sparse reconstruction algorithms to obtain the optimum solution (e.g., \emph{Basis Pursuit}, \emph{Orthogonal Matching Pursuit}, etc. (see \cite{Candes, Tropp} and references therein)).

In \cite{Tian2}, frequency sparsity is exploited for MB spectrum sensing. Nevertheless, the spectrum occupancy is measured based on the PSD which is still prone to estimation errors due to noise uncertainty. To overcome this, the authors in \cite{Tian3} exploit the cyclic-sparsity where a sub-Nyquist cyclostationarity detector is used. It is shown that not only the frequency spectrum is sparse, but also the two-dimensional (2D) cyclic spectrum is sparse too since some cyclic frequencies may not be occupied by any PUs. Thus, we can solely reconstruct the cyclic spectrum using the sub-Nyquist compressive samples without the need to recover the original signal or its frequency response. In addition, the proposed cyclic-CS (C-CS) can be used to estimate the PSD at lower complexity if the PU signals are assumed to be stationary. The advantage of C-CS is that it is robust to noise uncertainty since noise is non-cyclic (i.e. it does not appear at when the cyclic frequency $\alpha \neq 0$).

\begin{figure}[!b]
\centerline{\subfigure[]{\includegraphics[width=3.5in]{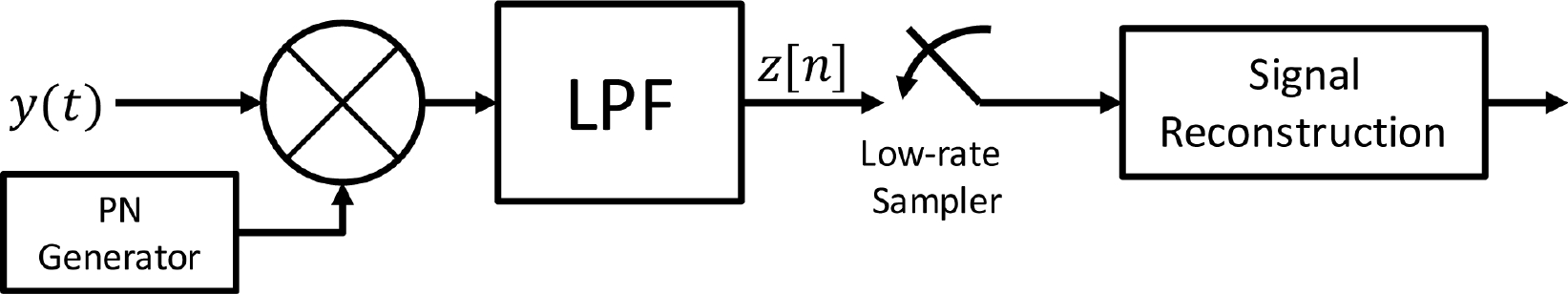}
\label{fig:Hatta6a}}}
\centerline{\subfigure[]{\includegraphics[width=3.5in]{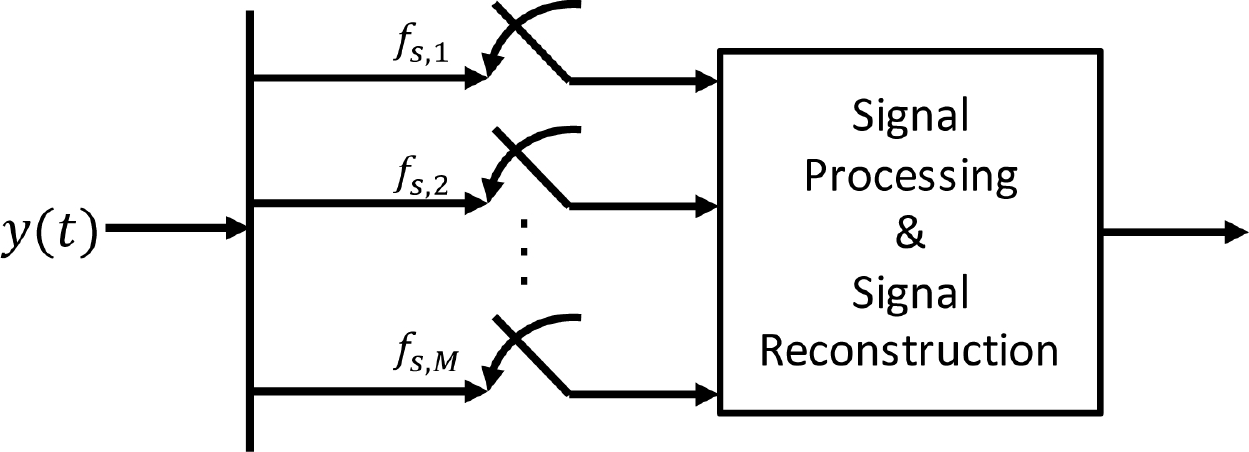}
\label{fig:Hatta6b}}}
\caption{(a) The analog-to-information converter; (b) The multi-rate sub-Nyquist sampling system ($t$ denotes continues time domain).}
\label{fig:Hatta6}
\end{figure}

The previous techniques assume that the signals are discrete. For analog signals, an analog-to-information (AIC) converter is used (also known as the \emph{random demodulator}) \cite{Kirolos1, Tropp}. The AIC basically extracts the information of the signal, and because the information rate of a sparse signal is less than its Nyquist rate, AIC promises to reduce the sampling burden. The structure of the AIC is shown in Fig. \ref{fig:Hatta6a}. The received signal is modulated by a pseudo-random number (PN) generator to spread the frequency content of the signal so that it is not destroyed by the low-pass filter (LPF). The signal is then sampled at a lower rate using any conventional Analog-to-Digital convertor (ADC). Then, using the appropriate CS algorithms, the signal can be successfully reconstructed from these partial measurements. An extension to this is presented in \cite{Polo2} where multiple SUs cooperate to improve the detection reliability. The drawback of this converter is that the PN generator is required for compression, and thus in order to exploit the spatial diversity in CRNs, each SU must have a separate compression device. In other words, synchronization among the SUs is required because asynchronized $\mathbf{\Phi}$ may degrade the spectrum reconstruction. To alleviate this deficiency, parallel sampling channels could be used. For instance, in \cite{Sun1}, the authors propose a multi-rate asynchronous sub-Nyquist sampling (MASS) system as shown in Fig. \ref{fig:Hatta6b}. That is, the system consists of $M$ sub-Nyquist sampling branches with each having a different low sampling rate, $f_{s}$. This structure does not require synchronization for generating $\mathbf{\Phi}$, has higher energy efficiency, and better data compression capability compared to AIC \cite{Sun1}.

To summarize, the CS is expected to significantly reduce the sampling rate, and hence the stringent requirements of ADC and receiver front-end might be relaxed. Nevertheless, there are some challenges to be further studied. For instance, careful analysis must be done regarding the SNR because it is expected to be degraded when the signal's information is compressed into smaller number of samples. Also, nonidealities in hardware will introduce noise to these measurements (e.g., the jitter in the PN generator, mixer's non-linearity, etc.) \cite{Kirolos2}. In addition, the reconstruction process is highly nonlinear unlike the Shannon sampling procedure, where the signal can be linearly reconstructed from its samples \cite{Tropp}. This means that to reduce the burden on the ADC, we need to pay it off with more complex signal processing (a tradeoff between software and hardware). In addition, all the aforementioned techniques assume that the sparsity basis is known. This opens a future research direction for robust CS with unknown basis. Finally, we have mentioned earlier that CS is a good candidate due to the spectrum sparsity. However, CRNs are expected to utilize these unoccupied bands, and hence upon their successful implementation, it is expected that the occupancy rate becomes higher (i.e. the wideband spectrum becomes less sparse), and thus CS loses its capability to compress the signals. In other words, CS may provide a solution to the spectrum scarcity problem, and ironically, this solution would hinder implementing CS afterwards! In general, CS is not effective to sense moderately or densely occupied spectrum.

\subsection{Angle-Based Sensing}
Most common spectrum sensing techniques are used to exploit the available opportunities in time, frequency, or space. That is, not all subchannels are occupied at the same time (opportunity in frequency domain), not all of them are permanently occupied (opportunity in time domain), and due the propagation losses in the wireless channels, the same channels may be reused in different geographical regions (opportunity in space). In \cite{Yucek}, a \emph{multi-dimensional} opportunistic access is outlined where new dimensions could be exploited such as the direction of arrival (DOA) domain (due to advancements in multi antenna technologies such as multiple-input multiple-output (MIMO) systems and beamforming).

The basic principle is that if the SU has a knowledge of the azimuth angle of the PUs, then when the PU transmits in a certain direction, the SU can simultaneously transmit in another direction, over the same band and geographical area. DOA-based Spectrum sensing is investigated for SB-CR in \cite{Jingjing}, and it is extended for MB-CR in \cite{Mahram}. DOA-based sensing can estimate the location of the occupied subbands by estimating the PU signals' direction of arrival. The drawback is that the SU must be equipped with a multi-antenna receiver where array processing becomes essential.

\subsection{Blind Sensing}
Blind sensing (BS) refers to the sensing problem when the structure of the received signal is unknown at the SU side. The energy detector can be considered, to a certain extent, as a blind detector since it does not require any prior knowledge of the PU signal, yet it still requires a good estimation of the noise variance. A more robust blind detector is presented in \cite{Luan} where a blind MJD exploits the eigenvalues of the covariance matrix of the received signal without estimating the noise variance.
For compressive sensing, the authors in \cite{Gleichman} propose a blind compressive sensing (BCS) algorithm where the sparsity basis, $\boldsymbol\Psi$, is unknown. They have shown that if the bases are orthogonal, the BCS algorithm performs very well. A multiband blind reconstruction of analog compressed signals is also investigated in \cite{Mishali}.

\subsection{Comparison}
Each multiband sensing technique poses some advantages and disadvantages as shown in Table \ref{tab:MultiBandDetectors}. Typically, serial spectrum sensing is relatively simple to implement. However, it is relatively slow, and this is undesirable especially when we have many subchannels. Some techniques have been proposed to make such algorithms faster, such as two-stage sensing, which requires additional components, and SPRT-based sensing, which has some practical disadvantages \cite[ch.~III]{Poor}. For example, SPRT techniques have a poor performance when the collected samples are correlated. Also, the number of required samples is a random variable and usually unbounded. Thus, some truncation techniques are necessary to limit the number of collected samples. In addition, SPRT techniques usually require prior knowledge of the PU signals, which is not easy to obtain.

Parallel sensing promises to provide faster detection, yet this is at the expense of more RF components at the SU receiver as well as more complex processing. MJD has become a benchmark due to its relative simplicity in terms of implementation (it basically consists of multiple energy detectors). However, optimizing the detection thresholds, $\boldsymbol{\lambda}$, is usually complex, and thus affordable suboptimal algorithms are preferred \cite{Beaulieu2}. Also, using a dynamic spectrum sensing time as in MSJD helps provide faster sensing.

Wavelet sensing is preferred when we want to estimate the number of subchannels and their corresponding boundaries or carrier frequencies. Compressive sensing promises to reduce the stringent sampling requirements, and angle-based sensing exploits a new dimension for opportunistic access. However, wavelet algorithms are susceptible to false edges, and even though several upgrades including WMM, WMP, and WMS are proposed to alleviate this limitation, they incur additional computational complexity. Similarly, compressive sensing requires knowledge of the sparsity basis matrix, and it depends on the fact that the spectrum of interest is sparse in some domain. Angle-based sensing requires array processing and multiple antennas that usually require larger receivers to be properly accommodated. Finally, blind sensing is practical since, in general, we do not have prior knowledge of the PUs. Unfortunately, the research on blind detectors for MB-CRNs is still limited.

\begin{table*}[!t]
\renewcommand{\arraystretch}{1.3}
\caption{A comparison between the common multiband spectrum sensing techniques}
\label{tab:MultiBandDetectors}
\small
\begin{center}
\begin{tabu}{l|[1pt]l|[1pt]l|[1pt]l}
\tabucline[1pt]{-}
\bfseries{Category}                             &   \bfseries{Sensing Algorithm}        &\bfseries{Advantages}              &\bfseries{Drawbacks}       \\\tabucline[1pt]{-}
\multirow{5}{2.5cm}{\bfseries{Serial-Based Detectors}}
            &   Reconfigurable BPF      &Simple                                       &High sampling rates; slow processing           \\\cline{2-4}
            &   Tunable Oscillator	    &Reduce the sampling rate	                  &Slow processing due to sweeping      \\\cline{2-4}
            &   Two-stage sensing	    &Faster sensing and improved detection        &Complex and expensive     \\\cline{2-4}
            &   SPRTs                   &Faster sensing	                              &They have practical challenges    \\\cline{2-4}
            &   Agile-based detector	&Faster sensing and improved detection	      &Complex and expensive     \\\tabucline[1pt]{-}

\multirow{2}{2.5cm}{\bfseries{Parallel-Based Detectors}}
            &   Filter Bank                         &Relatively simple                      &Expensive                      \\\cline{2-4}
            &   Frequency-based sensing             &Improved detection                   &Complex and expensive            \\\tabucline[1pt]{-}

\multirow{4}{2.5cm}{\bfseries{Wideband-Based Detectors}}
            &   Wavelet Sensing                   &Used when boundaries are unknown         &May detect false edges \\\cline{2-4}

            &Compressive Sensing                    &Tangibly reduce the sampling rate      &Requires knowledge of $\boldsymbol\Psi$  \\\cline{2-4}
            &Angle-Base Sensing                     &Exploits new dimension for spectrum access                &Requires multi-antenna system          \\\cline{2-4}
            &Blind Sensing                          &Good in the absence of prior knowledge       &Requires good estimation techniques
\\

\tabucline[1pt]{-}
\end{tabu}
\end{center}
\end{table*}

\subsection{Practical Implementation}
Multiband sensing techniques have truly inspired the research community in the last few years. Even though most of the published work is theory-oriented, few contributions have applied these underlying concepts into practical implementation for cognitive radio. In this section, we highlight some of the experiments and implementations for single-band and multiband sensing. The interested reader may refer to \cite{Cabric4} for a comprehensive survey of the main platforms and testbeds relating to cognitive radio.

Early work has focused on hardware implementation for single-band detectors. An experiment is implemented to study the performance of energy detection in real-environments in \cite{Cabric3}. Another experimental study is implemented in \cite{Cabric2} to evaluate the performance of feature detection using pilot signals. In these experiments, it is shown that the performance of the energy detector is bounded by noise power estimation, whereas feature detectors' performance is bounded by synchronization accuracy. Also, the performance can be enhanced if multiple users cooperate and share their local sensing results.

In \cite{Tkachenko}, an experiment shows that feature detectors based on cyclostationarity are susceptible to sampling clock offset. Similarly, the authors in \cite{Yonghong} show that imperfect estimation of the transmitter's cyclic frequency (also known as \emph{cyclic frequency offset}) results in an SNR wall (where detection becomes impossible). The authors in \cite{Cabric5} show that this wall can be overcome by splitting the observed samples into multiple blocks. In particular, a \emph{multi-frame} test statistic is proposed, and the results are validated via actual implementation.

The impact of third-order nonlinearities on the wideband receiver front-end is analytically analyzed for wideband energy and cyclostationarity detectors in \cite{Cabric7}. It is emphasized that adaptive interference cancellation algorithms are necessary to overcome these imperfections, which arise in practical implementations. In \cite{Kosunen}, implementations of various feature detectors based on autocorrelation and cyclostationarity features are presented.

One of the parameters that stand out in practical implementation is power consumption. Thus, many researchers and engineers have been motivated to design  chips and spectrum sensing processors that are power-efficient. For instance, the authors in \cite{Cabric6} test a wideband spectrum sensing processor, which is capable to sense a 200MHz channel bandwidth with very high efficiency. Hardware implementations of the MJD are also presented in \cite{Srinu, Kitsunezuka}.

The sampling rate imposed on the ADC is of great importance, and thus several works have focused on the implementation of compressive sensing for cognitive radio applications. The reader can refer to \cite{Duarte2} for a great connection between the theory of CS and its implementation. A data acquisition front-end based on CS is developed in \cite{Xi} to reduce the sampling rate requirement. It is shown that PN generators impose a major difficulty due to their power consumption, and it is suggested to use CMOS technology to mitigate this issue. Furthermore, a test-platform is developed for edge-detection using wavelet-based algorithms in \cite{Chantaraskul}. Similarly, the authors in \cite{Rebeiz} develop an energy-efficient processor to perform blind-based sensing.

\section{Cooperative Communications in Multiband Cognitive Radio Networks}
Significant breakthroughs have been accomplished in cooperative communication networks \cite{Khaled1}, yet most of them assume a single channel, and the work on cooperating SUs over multiple bands is still limited.

A cooperative compressive sensing based on hard-combining is investigated in \cite{Tian4} where each SU individually performs compressive sensing, then they share binary decisions with one another. A more practical scenario is studied in \cite{Tian5} where the channel-state information (CSI) is assumed to be unknown at the SU side. A soft-combining cooperative network is proposed for CS of analog signals in \cite{Polo2}. Cooperation in CS provides two advantages. First, the detection performance is enhanced due to the spatial diversity, and second, as the number of cooperating SUs increases, we can increase the compression ratio (i.e. reduce the sampling rate further) without performance degradation.
Cooperation is also investigated with respect to the network's throughput \cite{Quan1, Quan3, Fan1, Fan2}. The MJD framework is extended to multiple SUs where a \emph{spatial-spectral joint detector} is proposed in \cite{Quan1, Quan3}. In this detector, the idea is to linearly combine the soft decisions made by the individual SUs such that a joint optimization of the detection thresholds and the weighting coefficients, as expressed in (\ref{eq:SoftCombining}), is implemented. It is demonstrated that cooperation significantly improves the throughput compared to a single MJD.

Other works include \cite{Derakhtian, Song, Kim1}. In \cite{Derakhtian}, several multiband detectors based on the generalized LRT (GLRT) are investigated under different fading channels. In \cite{Song}, the impact of noise uncertainty on cooperative sensing is studied. The authors propose an algorithm where the SU first performs multiband spectrum sensing. For subchannels that have uncertain noise power estimation, the SU cooperates with the neighboring SUs to check their decision of these uncertain channels. In \cite{Kim1}, sequential cooperation is investigated. That is, for a certain channel, the fusion center will sequentially collect one-bit decisions from the SUs until a decision is made. It is shown that subchannels with uncertain SU decisions require more cooperating SUs to mitigate the uncertainties.

\begin{figure}[!b]
\centerline{\subfigure[]{\includegraphics[width=3.5in]{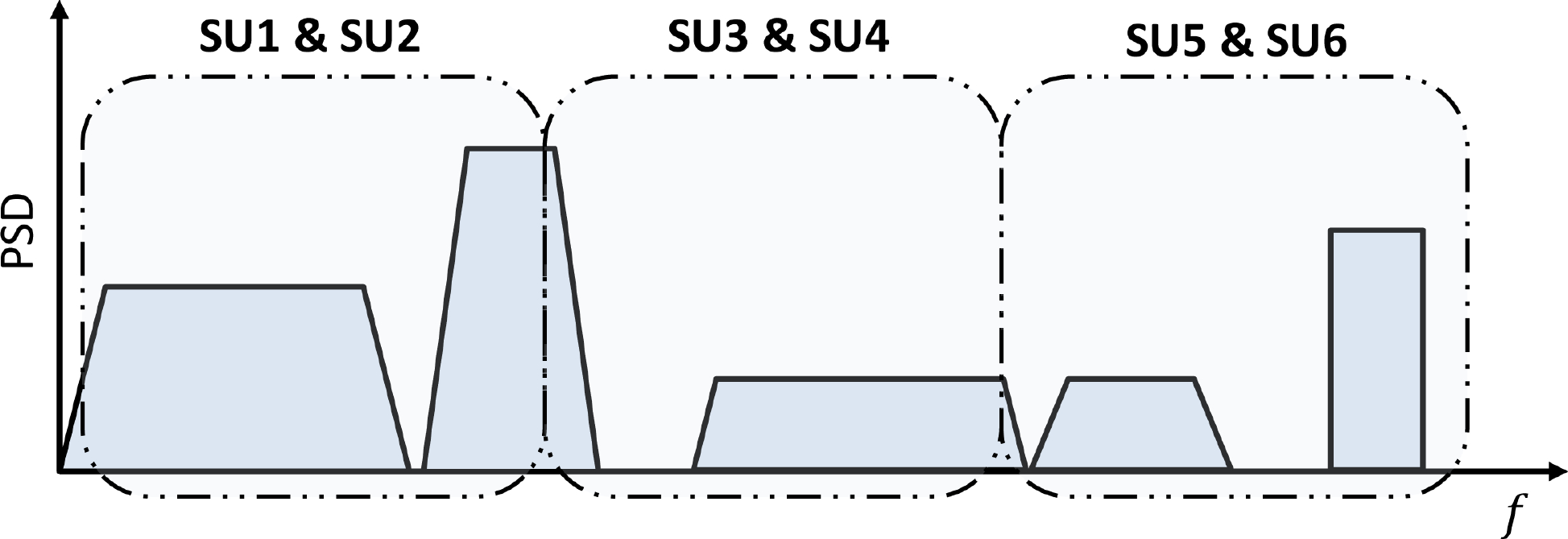}
\label{fig:Hatta7a}}}
\centerline{\subfigure[]{\includegraphics[width=3.5in]{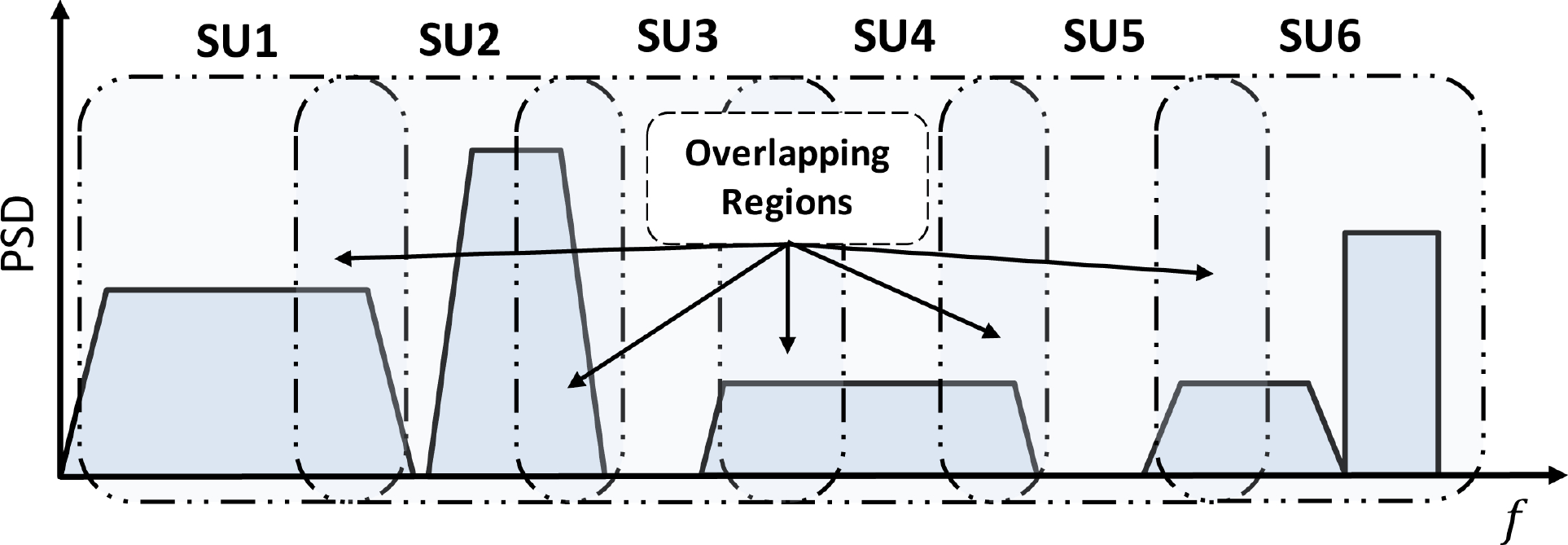}
\label{fig:Hatta7b}}}
\caption{(a) A uniform diversity of 2; (b) Non-uniform diversity.}
\label{fig:Hatta7}
\end{figure}

The previous contributions assume that each SU senses the entire spectrum before cooperation. This hefty load and the load of cooperating over the multiple bands make the implementation implausible even if hard combining is used since each SU must represent the $M$ subchannels by an $M\times1$ binary vector. Thus, instead of making the maximum gains of both paradigms, one can restrict to a tradeoff that makes network implementation more feasible where each SU senses a subset of $M$ subchannels. We shall refer to this paradigm as \emph{cooperative multiband cognitive radio networks}.
Consider Fig. \ref{fig:Hatta7} where each SU senses a subset of these bands such that the entire spectrum is sensed by all of them together. Since we have 6 SUs, the maximum possible spatial diversity is 6, which is attained when each SU senses all of the $M$ subchannels. This is very demanding in terms of the sampling requirements. To reduce these requirements, compressive sampling may be implemented. Alternatively, each SU can monitor a subset of these channels to reduce the sampling requirements. In Fig. \ref{fig:Hatta7a}, uniform diversity is achieved, with a value of 2. That is, each subchannel is being sensed by two SUs. Another approach is to use non-uniform diversity where the number of SUs monitoring a certain band depends on several factors. For example, if the band has higher priority (e.g., bands for public safety), or has intense PUs activities (e.g., cellular bands), then more SUs may be allocated to sense such band to improve detection reliability. This is shown by the \emph{overlapping regions} in Fig. \ref{fig:Hatta7b}.
Such paradigm involves a fundamental tradeoff between spatial diversity and sampling requirements. This tradeoff will be explained in Section VI.

\section{Performance Measures}
In this section, we categorize the performance measures into two broad categories: spectrum sensing performance measures, and network's throughput.

\subsection{Spectrum Sensing Performance Measures}
The \emph{receiver operating characteristics} (ROC) is probably the most common performance metric in spectrum sensing. It is a plot of the probability of detection, $P_D$, versus the probability of false alarm, $P_{FA}$. We will revise these definitions for single-band spectrum sensing with and without cooperation.

\subsubsection{Single Band}
\begin{figure}[!t]
\centering
\includegraphics[width=3.5in]{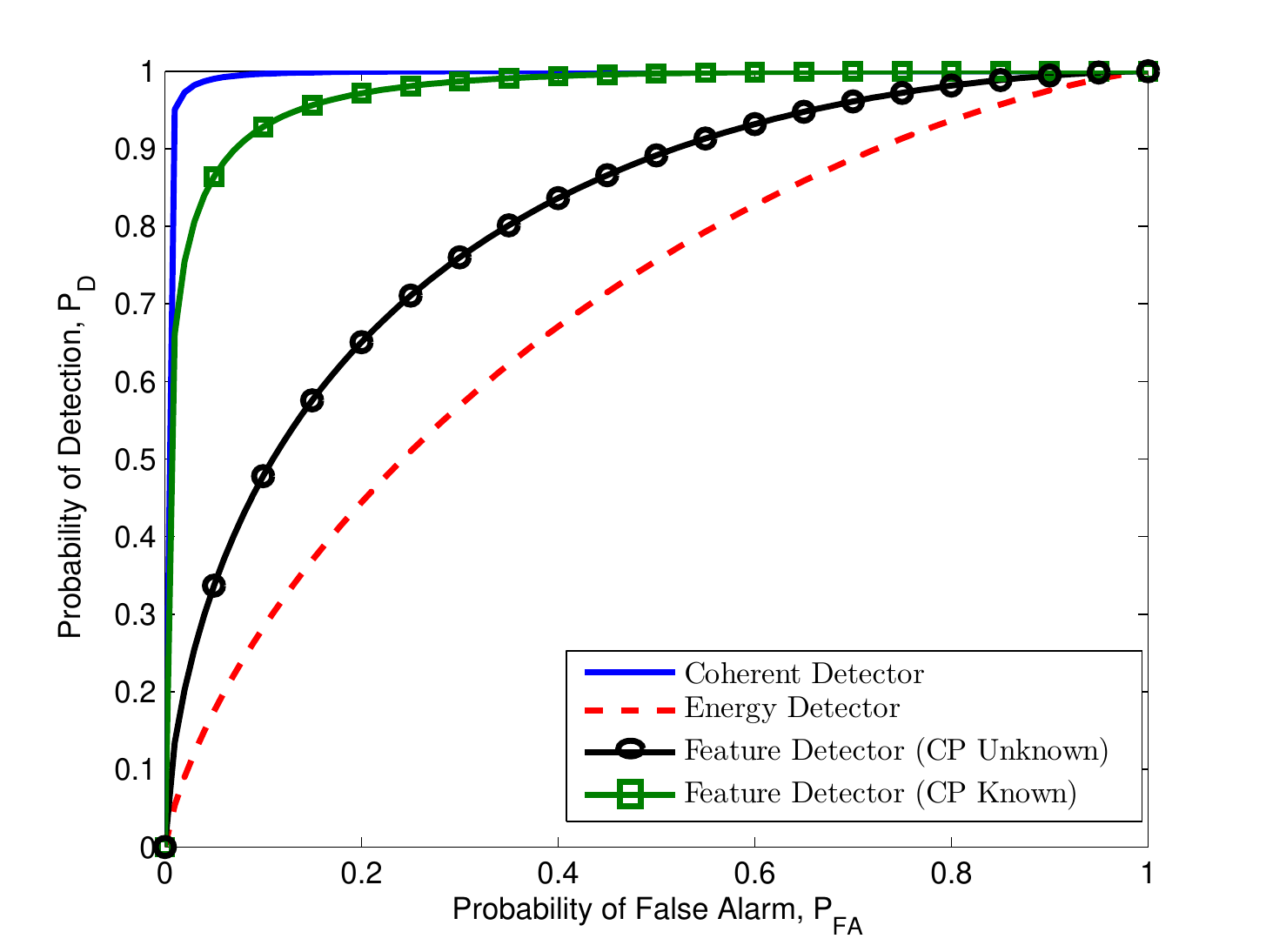}
\caption{The ROC curves of three different single-band detectors.}
\label{fig:Hatta8}
\end{figure}

For single-band CR, $P_D$ is simply the probability that the SU correctly detects the PU when it is present in a given band. Hence, for the given test in (\ref{eq:LRT}), $P_D$ is expressed as
\begin{equation}
\label{eq:DetectionProbability}
P_D=\text{Pr}(T(\mathbf{y})>\lambda |\mathcal{H}_1).
\end{equation}
In contrary, $P_{FA}$ is the probability that the SU incorrectly decides the presence of PUs albeit they are actually idle. This is expressed as
\begin{equation}
\label{eq:FalseAlarmProbabiltiy}
P_{FA}=\text{Pr}(T(\mathbf{y})>\lambda |\mathcal{H}_0).
\end{equation}
It is desirable to have higher probability of detection and lower probability of false alarm. The former guarantees minimal interference with the PU, and the latter guarantees throughput improvements for the secondary users. Nevertheless, a tradeoff between these two is inevitable. For example, if the SU has a full knowledge of the PU transmitted signal, then $\mathbf{x}$ is deterministic, and hence for the given model in (\ref{eq:SingleBandSpectrumSensingProblem}), we have
\begin{equation}
\begin{aligned}
\mathcal{H}_0:  &~  T(\mathbf{y})\sim \mathcal{N}(0,||\mathbf{x}||^2\sigma^2)\\
\mathcal{H}_1:  &~  T(\mathbf{y})\sim \mathcal{N}(||\mathbf{x}||^2,||\mathbf{x}||^2\sigma^2).
\end{aligned}
\end{equation}
With direct computations of (\ref{eq:FalseAlarmProbabiltiy}) and (\ref{eq:DetectionProbability}), we have \cite[ch.~III]{Poor}
\begin{equation}
P_{D}   =  Q\bigg(Q^{-1}\big(P_{FA}\big)-\sqrt{N\gamma}\bigg),
\end{equation}
where $\gamma=||\mathbf{x}||^2/\sigma^2$ denotes the SNR, $Q(.)$ is the complementary distribution function of the standard Gaussian, and $Q^{-1}(.)$ is its inverse. Alternatively, when the SU does not have prior knowledge about $\mathbf{x}$, we can assume that $\mathbf{x}\sim\mathcal{N}(0,\sigma_s^2\mathbf{I})$, and use the energy detector. Therefore,
\begin{equation}
\begin{aligned}
\mathcal{H}_0:  &~  T(\mathbf{y})\sim \mathcal{X}_N^2\\
\mathcal{H}_1:  &~  T(\mathbf{y})\sim \frac{||\mathbf{x}||^2+\sigma^2}{\sigma^2}\mathcal{X}_N^2,
\end{aligned}
\end{equation}
where $\mathcal{X}_N^2$ denotes a central chi-square distribution with $N$ degrees of freedom. Using the central limit theorem, It can be shown that \cite{Liang}
\begin{equation}
P_{D}   =  Q\bigg(\frac{1}{\sqrt{2\gamma+1}}\Big(Q^{-1}\big(P_{FA}\big)-\sqrt{N}\gamma\Big)\bigg).
\end{equation}
In Fig. \ref{fig:Hatta8}, we illustrate the performance of three main detectors: The coherent detector, the energy detector, and a feature detector that exploits the 2nd order statistics of an orthogonal frequency division multiplexing (OFDM) signal under two different scenarios. In the first one, the SU has knowledge about the number of useful symbols in the OFDM block, and in the 2nd scenario, it has additional knowledge about the duration of the cyclic prefix (CP) (see \cite{Chaudhari} for derivations). We assume that $SNR=-15$dB, and the number of observed blocks $N=500$. We observe the following. First, the best performance is attained by the coherent detector since the SU is assumed to have full knowledge of $\mathbf{x}$. Second, the feature detector has an excellent performance when more features are exploited such as the duration of the CP. Third, the energy detector has a poor performance at the low SNR region\footnote{SUs must be able to detect PU signals as low as -114dBm \cite{FCC3}. Roughly speaking, in the low SNR region, we have $SNR\leq-10$ dB.}. This region is important because the PU signals might be weak at the SU receiver, and to reliably detect them, it must be equipped with a detector that performs well under low SNR conditions \cite{Cabric1}.

\subsubsection{Cooperative Spectrum Sensing}
\begin{figure}[!t]
\centering
\includegraphics[width=3.5in]{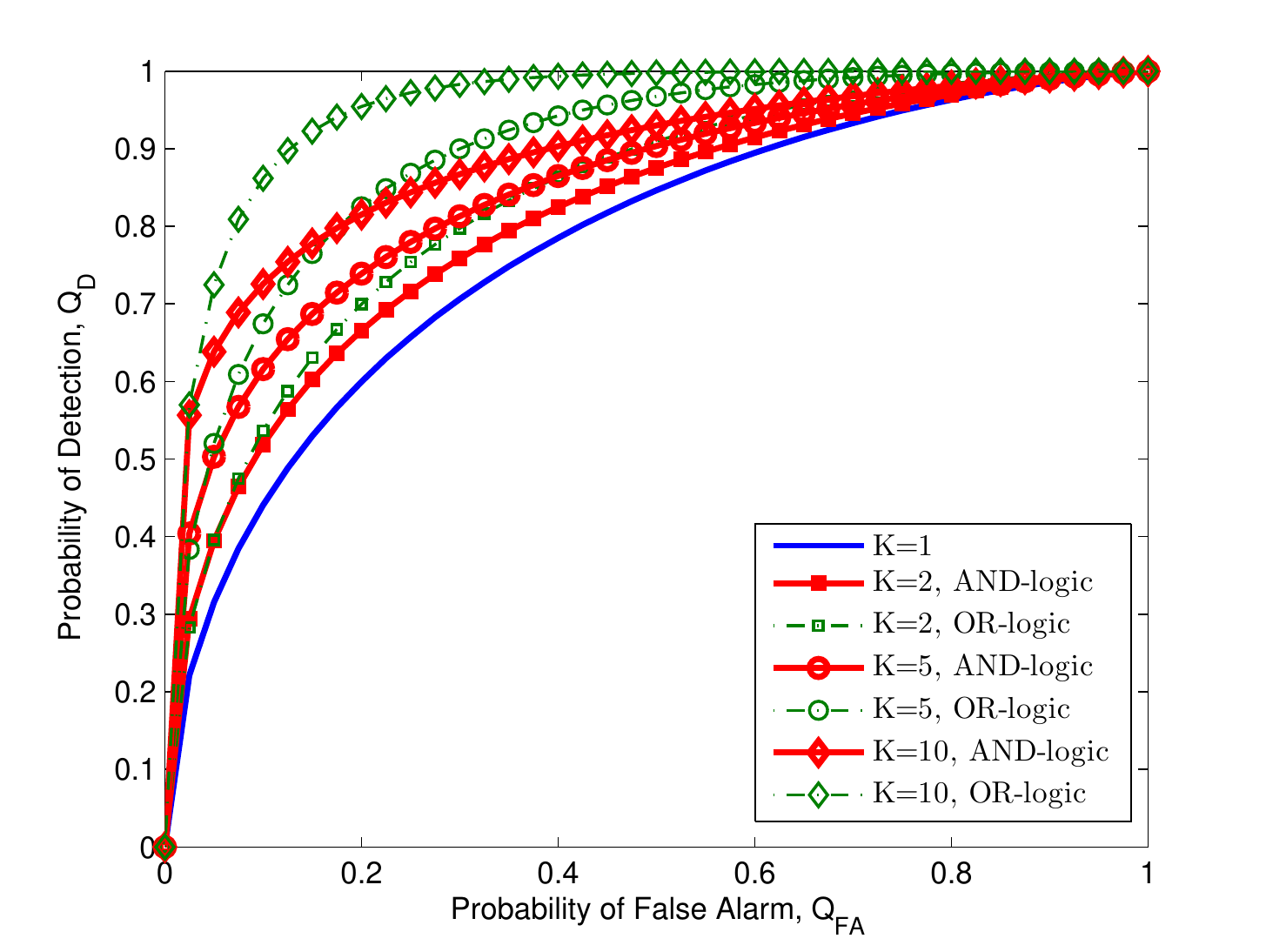}
\caption{The ROC curves with variations of number of cooperating SUs.}
\label{fig:Hatta9}
\end{figure}

In cooperative spectrum sensing, let $P_D^{(i)}$ and $P_{FA}^{(i)}$ denote the detection and false alarm probabilities of the $i$-th SU, respectively. Then, in hard combining, if we have a  $k$ out of $K$ rule, then the overall detection and false alarm probabilities are, respectively,
\begin{equation}
Q_D     =   \sum_{q=k}^{K}   {{K}\choose{q}} \bigg\{ \prod_{i=1}^q P_D^{(i)} \times \prod_{j=1}^{K-q} (1-P_D^{(j)}) \bigg\}
\end{equation}
\begin{equation}
Q_{FA}     =   \sum_{q=k}^{K}   {{K}\choose{q}} \bigg\{ \prod_{i=1}^q P_{FA}^{(i)} \times \prod_{j=1}^{K-q} (1-P_{FA}^{(j)}) \bigg\},
\end{equation}
Fig. \ref{fig:Hatta9} shows the ROC of an energy detector with different number of cooperating SUs. We assume that all SUs have identical $P_{FA}$ and $P_D$ performance, $SNR=-10$dB, and $N=125$. It is observed that increasing the number of cooperating SUs improves the performance. Also, the OR-logic rule provides a better performance compared to the AND-logic rule.

For soft combining, the detection and false alarm probabilities must be derived explicitly for a given model. For instance, the probability of detection of $\mathcal{T}$ in (\ref{eq:SoftCombining}) is
\begin{equation}
P_D =   \text{Pr}\Big(\sum_{k=1}^{K}  c_k T_k(\mathbf{y})>\lambda|\mathcal{H}_1\Big).
\end{equation}
Thus, once we find the probability distribution of $\mathcal{T}$, we can use the above expression to find $P_D$, and a similar procedure is required to find $P_{FA}$.

\subsubsection{Multiband Cognitive Radio}
Unlike single-band sensing, there is no unified definition for the detection and false alarm probabilities when multiple bands are considered. Intuitively, one can calculate these probabilities for each subband individually. That is, we compute $P_{D,m}$ for $m=1,2,\ldots,M$. Another naive approach is to average the performance of all the bands together. That is,
\begin{equation}
\label{eq:AveragePD}
P^M_D     =       \frac{1}{M} \sum_{m=1}^{M} P_{D,m}.
\end{equation}
Nevertheless, averaging may lead to incorrect insights especially when there are outliers (i.e if a single channel has a very low detection probability and the other channels have very good performance, the average detection probability is going to be low). A more robust approach could be to use a normalized weighting average such as
\begin{equation}
\label{eq:WeightAveragePD}
P^M_D     =      \sum_{m=1}^{M} a_m P_{D,m},
\end{equation}
where $a_m$ is the weighting coefficient of subband $m$ subject to $\sum_{m=1}^{M}a_m=1$. These weights could reflect the sensitivity or the importance of a channel. For instance, if a channel does not require tight interference requirements, one can assign a low weight to that channel such that the performance report of that channel does not have significant impact on the overall performance. Clearly, when the weights are equal, (\ref{eq:WeightAveragePD}) reduces to (\ref{eq:AveragePD}).

In \cite{Zeng2}, another definition of these probabilities is proposed such that
\begin{equation}
\begin{aligned}
P_D^M     &=   p(\text{at least one band detected as occupied}|\mathcal{H}_1^M)   \\
P^M_{FA}  &=   p(\text{at least one band detected as occupied}|\mathcal{H}_0^M),
\end{aligned}
\end{equation}
where $\mathcal{H}_1^M$ and $\mathcal{H}_0^M$ denote the events that all channels are busy and all channels are vacant, respectively. In \cite{YanXin}, the false alarm probability is defined as the probability of deciding that all channels are busy even though there exists at least one vacant channel. If we let $\mathcal{H}_0^{(i)}$ denote that there exist at least $i$ free channels for $i=1,\ldots,M$. Then the probability of false alarm can be rewritten as $P_{FA}^{M}=p(\mathcal{H}_1^{M}|\bigcup_{i=1}^{M}\mathcal{H}_0^{(i)})$, which can be also expressed as
\begin{equation}
\label{ModifiedPf}
P_{FA}^{M}  =   \frac{\sum_{i=1}^{M}p(\mathcal{H}_1^{M}\cap\mathcal{H}_0^{(i)})}{p(\bigcup_{i=1}^{M}\mathcal{H}_0^{(i)})}.
\end{equation}

While the previous measures assume perfect knowledge of the subchannel boundaries, this is not the case in edge-detection (i.e. in wavelet detection) where the performance of estimating these edges is also necessary. The authors in \cite{ElKhamy} propose new measures to test the performance of the edge detection techniques. If we denote $N_B=M+1$ to be the actual number of subchannels' boundaries in a system, and $\hat{N}_B$ to be the number of boundaries that are correctly detected, then the probability of miss detecting an actual boundary is
\begin{equation}
P_{ME}  =   \frac{N_B-\hat{N}_B}{N_B},
\end{equation}
and the probability of detecting a false-edge is given by
\begin{equation}
P_{FE}  =   \frac{N_T-\hat{N}_B}{N_{FFT}-N_B},
\end{equation}
where $N_T$ is the total number of all detected edges including both the actual subchannels' edges and the false edges, and $N_{FFT}$ is the utilized FFT size. Clearly, increasing the size of FFT will improve the detection performance. Finally, the average edge detection error probability is defined as
\begin{equation}
P_E     =   \frac{P_{MR}+P_{FE}}{2}.
\end{equation}
Alternatively, in \cite{Zeng2}, the authors define the \emph{band occupancy error} (BOE) as
\begin{equation}
\label{eq:BOE}
\text{BOE}  =   \frac{\sum_{m=1}^{M} |\text{BOD}_a(m)-\text{BOD}_e(m)|^2}{\sum_{m=1}^{M} |\text{BOD}_a(m)|^2},
\end{equation}
where the subscripts $a$ and $e$ indicate, respectively, the actual and the estimated \emph{band occupancy degree} (BOD), which is expressed as
\begin{equation}
\label{eq:BOD}
\text{BOD}(m)\left\{
\begin{array}{ll}
\gamma_m,     &     \text{subchannel } m \text{ is occupied} \\
0,            &     \text{otherwise}
\end{array} \right.,
\end{equation}
where $\gamma_m$ is the SNR at band $m$. A very similar procedure is also proposed for individual subcarriers within each band. Note that a simpler BOD is achieved if we replace $\gamma_m$ with 1. That is, `0' and `1' represent an unoccupied and occupied subchannel, respectively.

In addition to the aforementioned metrics, several modified versions can be used such as the complementary ROC where we plot the probability of miss detection $P_M=1-P_D$ versus $P_{FA}$ or $Q_M=1-Q_D$ versus $Q_{FA}$ in case of using cooperative communications. Other intuitive performance metrics are based on plotting these probabilities against the SNR.

\subsection{Throughput Performance Measures}
One of the main motivations of MB-CRNs is that they potentially enhance the network's throughput, and perhaps guarantee some QoS provisioning for SUs. Therefore, throughput is an important performance measure in MB-CRNs.

\begin{figure}[!b]
\centering
\includegraphics[width=3.5in]{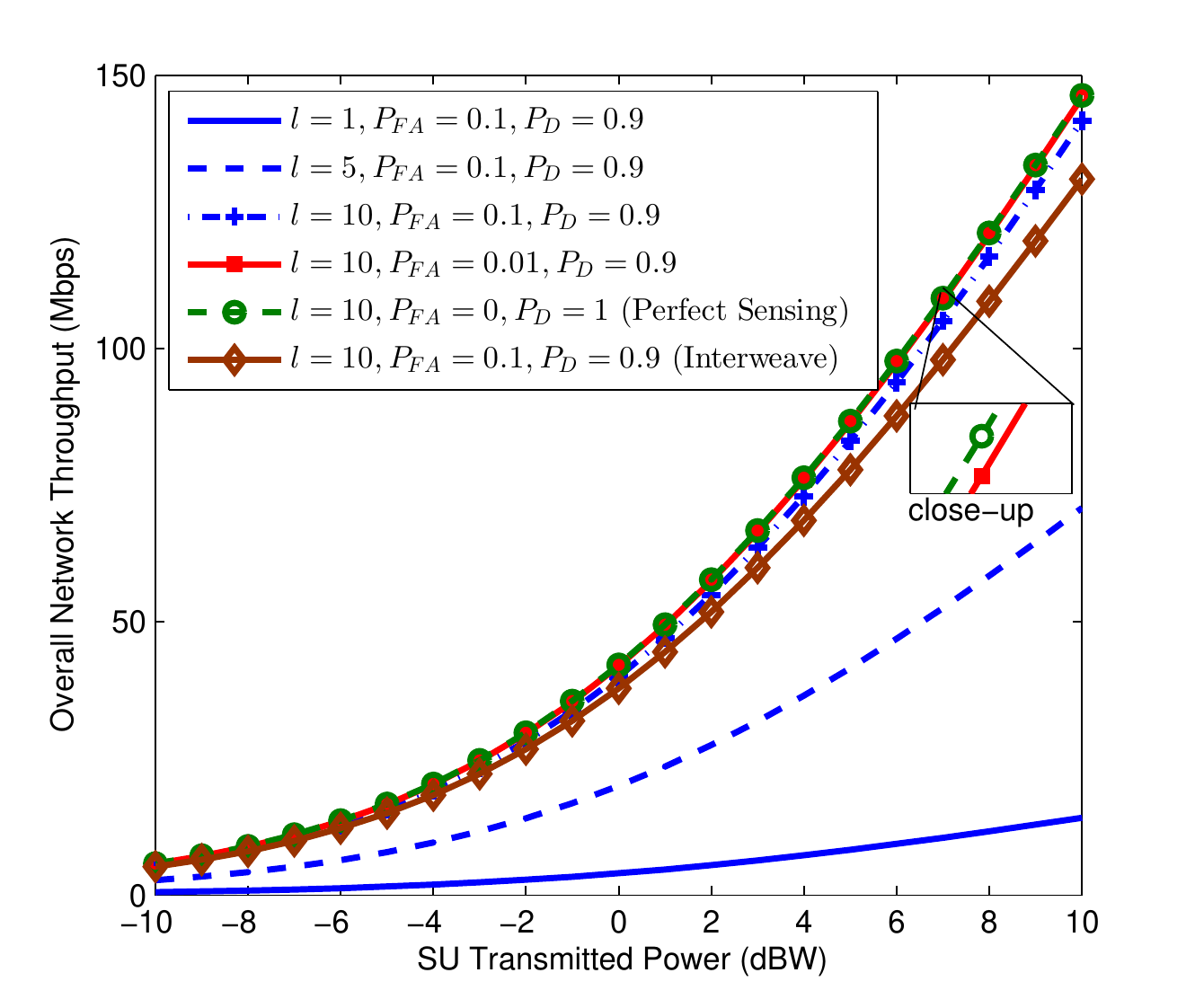}
\caption{The network's throughput with variations of SU transmitted powers under different number of accessed subchannels ($l=1,5,$ and $10$). Here, we assume that $p(H_{0,m})=1-p(H_{1,m})=0.7$ for $m=1,2,\ldots,l$.}
\label{fig:Hatta10}
\end{figure}

We consider a general transmission paradigm where the SU accesses the band whether the PU is absent (interweave paradigm) or the PU is present (underlay paradigm). In the latter, power adaptation is mandatory to protect the PUs. Such combination of the two is commonly known as \emph{sensing-based spectrum sharing} \cite{Xin}. Furthermore, we assume imperfect sensing. Under these two assumptions, there are four possible scenarios for SUs transmission:

\begin{itemize}
\item The SU correctly decides that the PU is absent, and thus it transmits at power $\mathcal{P}_s^{0}$ with probability $1-P_{FA,m}$.
\item The SU incorrectly decides that the PU is absent, and thus it transmits at power $\mathcal{P}_s^{0}$ with probability $1-P_{D,m}$.
\item The SU correctly decides that the PU is present, and thus it transmits at power $\mathcal{P}_s^{1}<P_s^{0}$ with probability  $P_{D,m}$.
\item The SU incorrectly decides that the PU is present, and thus it transmits at $\mathcal{P}_s^{1}$ with probability $P_{FA,m}$.
\end{itemize}
We denote $r_{ij}$ to be the transmission rate of the SU given that it decides $\mathcal{H}_i$ when $\mathcal{H}_j$ is the true hypothesis. Thus, we have \begin{equation}
\begin{aligned}
\label{eq:ShannonFormula}
r_{00}      &=   B \log_2 \Big(1+\frac{\mathcal{P}_s^{0}}{\sigma^2}\Big)    \\
r_{01}      &=   B \log_2 \Big(1+\frac{\mathcal{P}_s^{0}}{\mathcal{I}+\sigma^2}\Big)    \\
r_{11}      &=   B \log_2 \Big(1+\frac{\mathcal{P}_s^{1}}{\mathcal{I}+\sigma^2}\Big)    \\
r_{10}      &=   B \log_2 \Big(1+\frac{\mathcal{P}_s^{1}}{\sigma^2}\Big),
\end{aligned}
\end{equation}
where $B$ is the subchannel bandwidth, and $\mathcal{I}$ is the interference due to the PU transmission when it is present. If we assume that the SU accesses $l$ subchannels out of $M$, then the average throughput of the network is given by
\begin{equation}\tag{39}
\begin{aligned}
\label{eq:Throughput}
R       &=    \sum_{m=1}^{l}    p(\mathcal{H}_{0,m})[r_{00,m} (1-P_{FA,m}) ]+   r_{10,m} (P_{FA,m})] \\
        &+   p(\mathcal{H}_{1,m})[r_{01,m} (1-P_{D,m})+r_{11,m} (P_{D,m})],
\end{aligned}
\end{equation}
where $p(\mathcal{H}_{0,m})$ and $p(\mathcal{H}_{1,m})$ denote the probability that channel $m$ is idle and active, respectively. We remark the following. First, under perfect sensing conditions, $r_{01}$ and $r_{10}$ become 0. Also, under the interweave paradigm (i.e. the SU only transmits when the PU is absent), the contribution of $r_{11}$ and $r_{10}$ become 0. A simple simulation is shown in Fig. \ref{fig:Hatta10} to illustrate the throughput improvements using MB-CRNs.  The subchannels are assumed to have bandwidth of 6MHz each with the same SNR conditions. The power allocation is uniformly distributed among the $l$ channels, $\mathcal{P}_s^{1}=0.4\mathcal{P}_s^{0}$, $\sigma^2=1$, and $\mathcal{I}=-20$dBW. As Fig. \ref{fig:Hatta10} indicates, if the SU accesses more channels (i.e. $l$ increases), the throughput would increase as MB-CRN promises. Also, for a tighter $P_{FA}$, the throughput is further improved since the false alarm probability is less, and hence the data interruptions are less frequent. Finally, sensing-based spectrum sharing (or hybrid access) gives better throughput in comparison with interweave access because it allows the SU to coexist with the PU.

\section{Fundamental Limits and Tradeoffs}
There are several key design parameters in MB-CRNs that must be carefully investigated. The most common considerations are the sensing time, network's throughput, detection reliability, number of cooperating SUs, power control, and channel assignment.

In general, the design procedure is as follows. For a set of parameters, the designer wants to choose the best values that maximize some function such as the throughput, or minimize another function such as interference to PUs. Mathematically, this can formulated as an optimization problem, which can have the form \cite{Boyd}
\begin{equation}
\label{eq:Optimization}
\begin{aligned}
\text{maximize }     &   f(\mathbf{o})   \\
\text{subject to }   &   g_i(\mathbf{o})  \leq b_i,   &i=1,2,...,I,
\end{aligned}
\end{equation}
where $f(\mathbf{o})$ is the objective function, $\mathbf{o}$ is the optimization variable, $g_i(\mathbf{o})$ are constraint functions bounded by $b_i$. In this section, we will illustrate the design tradeoffs, and discuss some of the techniques that could provide improvements to these parameters.
Recall that we have $M$ subchannels, $K$ SUs, and each one of them observes $N$ samples.

\begin{figure}[!b]
\centerline{\subfigure[]{\includegraphics[width=3.5in]{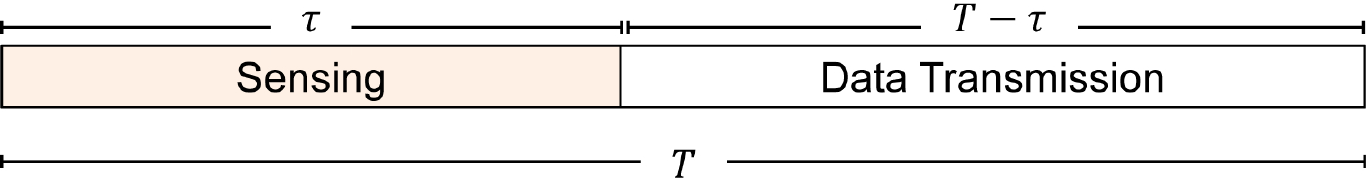}
\label{fig:Hatta11a}}}
\centerline{\subfigure[]{\includegraphics[width=3.5in]{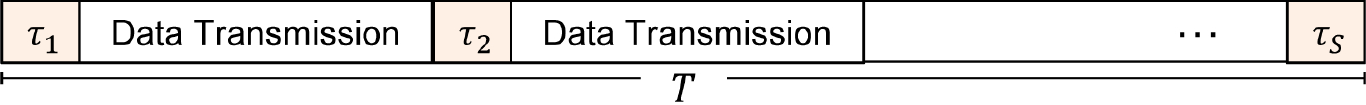}
\label{fig:Hatta11b}}}
\centerline{\subfigure[]{\includegraphics[width=3.5in]{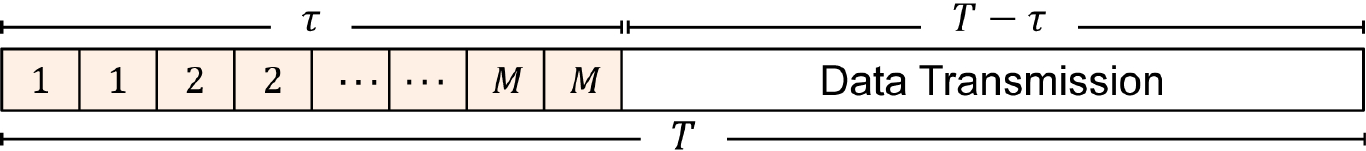}
\label{fig:Hatta11c}}}
\centerline{\subfigure[]{\includegraphics[width=3.5in]{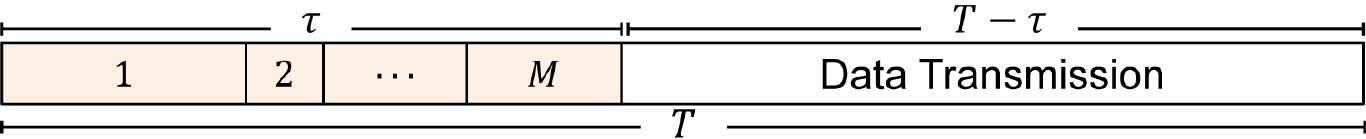}
\label{fig:Hatta11d}}}
\centerline{\subfigure[]{\includegraphics[width=3.5in]{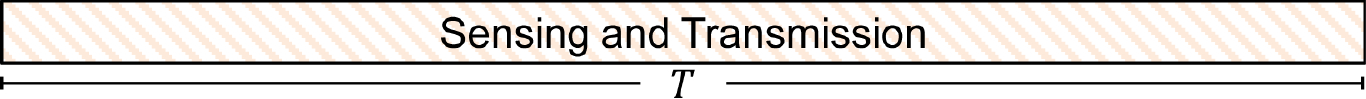}
\label{fig:Hatta11e}}}
\centerline{\subfigure[]{\includegraphics[width=3.5in]{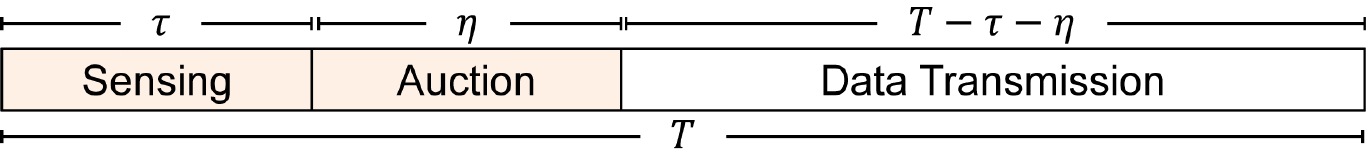}
\label{fig:Hatta11f}}}
\caption{(a) A conventional MAC frame in CR; (b) SB slotted frame; (c) MB slotted frame; (d) MB arbitrary-length slotted frame; (e) A novel frame where transmission and sensing occur simultaneously; (f) Auction based frame. }
\label{fig:Hatta11}
\end{figure}

\begin{figure}[!b]
\centerline{\subfigure[]{\includegraphics[width=3.5in]{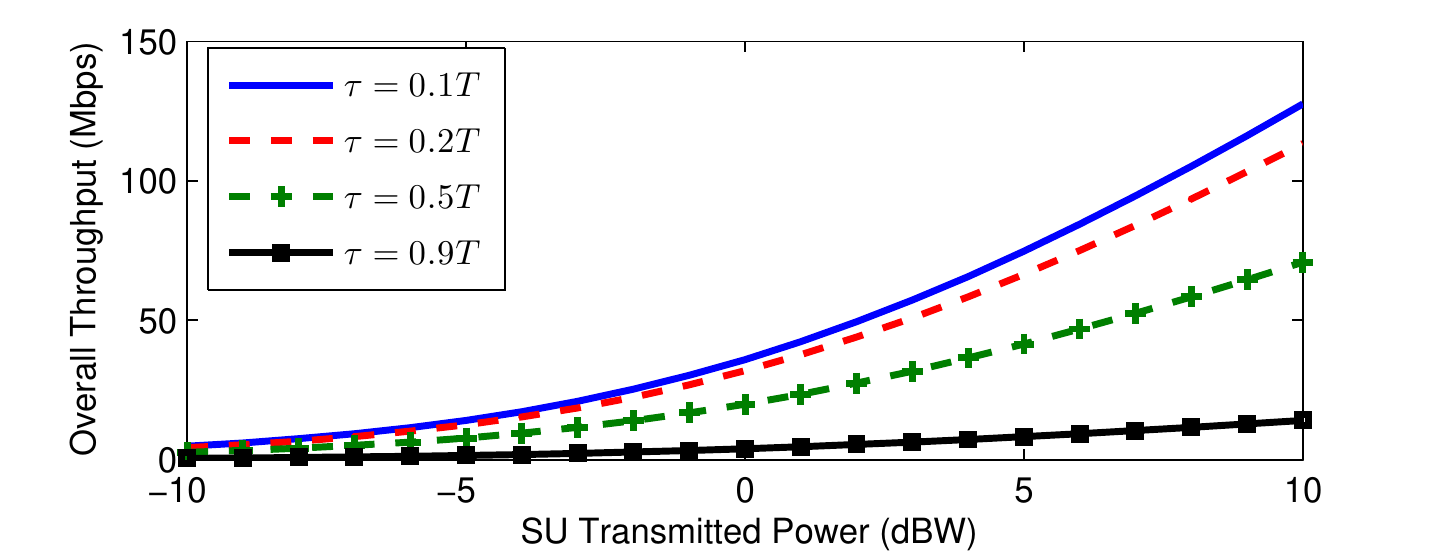}
\label{fig:Hatta12a}}}
\centerline{\subfigure[]{\includegraphics[width=3.5in]{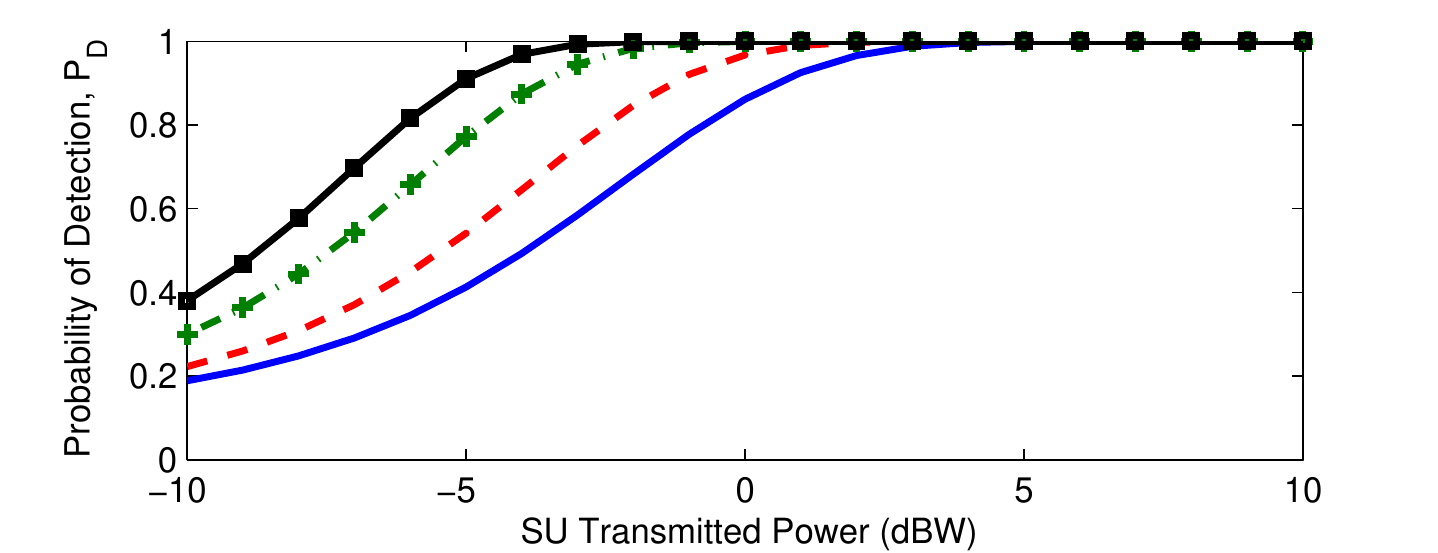}
\label{fig:Hatta12b}}}
\caption{Sensing time provides a tradeoff between throughput and detection reliability ($l=10, P_{FA}=0.1$). }
\label{fig:Hatta12}
\end{figure}

\subsection{Sensing Time Optimization}
One of the key parameters in spectrum sensing is the sensing duration, which strongly impacts the network's throughput. To illustrate this, Fig. \ref{fig:Hatta11a} shows a MAC frame structure that has been widely adopted for cognitive radio networks. The $T$ seconds frame consists of two slots: a sensing slot, $\tau$, and a transmission slot, $T-\tau$, if assuming the SU accesses the subchannel. Thus, when sensing duration is considered, the throughput becomes
\begin{equation}
\label{eq:ActualThroughput}
C=  \frac{T-\tau}{T} R.
\end{equation}
Clearly, increasing $\tau$ will shorten the transmission slot, and the throughput of the SU is reduced. This is shown in Fig. \ref{fig:Hatta12a}. However, longer sensing improves the detection performance since $N=\tau f_s$, where $f_s$ is the sampling frequency (i.e. the SU receiver collects more samples for its test statistics), and this is shown in Fig. \ref{fig:Hatta12b}, where we assume we are using an energy detector.
Liang et al. have  studied this tradeoff in \cite{Liang} for SB-CR.

Mathematically, the optimization problem can be written as
\begin{equation}
\label{eq:TimeOptimization}
\begin{aligned}
\max_{\tau}         \text{ }    &   C(\tau)                         \\
\text{s.t.}         \text{ }    &   P_D(\lambda)  \geq \beta;~     ~ 0<\tau<T,
\end{aligned}
\end{equation}
where $\beta$ is simply a constraint on $P_D$, and it is often known as the \emph{target probability}. Other variations arise in the literature where additional constraints are used. For example, an additional constraint on $P_{FA}$ could be added, or the detector's thresholds can be jointly optimized with sensing time (e.g., $\mathbf{o}=[\tau,\boldsymbol\lambda]$), which is the case in the MSJD detector. Next, we will discuss some of the techniques that can be used to balance the sensing time-throughput tradeoff.

\subsubsection{Different MAC Frame Structures}
The authors in \cite{Liang} have proposed a modified frame structure as shown in Fig. \ref{fig:Hatta11b} where the $\tau$ sensing duration is decomposed into $S$ equal sensing slots. It is shown that increasing the number of these slots would reduce the optimum sensing time, and thus the throughput is improved, and interestingly, the probability of false alarm is also reduced. This work is extended for MB serial sensing in \cite{Fan1} where the authors analyze two different frame structures: multiband slotted-frame where each subchannel is allocated a fixed number of slots, and multiband continues-frame, where each subchannel has an arbitrary-length duration bounded by the total sensing time as shown in Fig. \ref{fig:Hatta11c} (each subchannel here is allocated two slotted-frames) and Fig. \ref{fig:Hatta11d}, respectively. In the practical sense, the slotted-frame is easier to implement, yet it is demonstrated that solving for the optimum time is less complex if we use the continuous-frame. It is also demonstrated that increasing the number of channels for sensing improves the throughput.

In \cite{Stotas2}, a novel frame structure is proposed where sensing and transmission are carried out simultaneously, as shown in Fig. \ref{fig:Hatta11e}, using decoding in parallel with spectrum sensing. The basic principle is follows. The SU initially senses the spectrum, and starts transmitting when a band is vacant. However, it transmits both its data and the spectrum occupancy to the base station. The base station decodes the received signal, extracts the SU data, and uses the rest of the signal to analyze the spectrum occupancy using any of the spectrum sensing techniques. Since both decoding and sensing are done in parallel, the advantages are: longer sensing duration that enhances the detection performance, longer transmission rate that enhances the throughput, better PU protection since the sensing is always carried on, and finally sensing time optimization becomes unnecessary since it is done for the entire frame duration. The drawbacks, however, are twofold. First, it is demonstrated in \cite{Tang} that decoding errors compromise the performance. Thus, it is shown that there is a lifetime of which this novel frame is superior to the conventional frame, and beyond it, the novel frame loses its advantages due to the accumulation of decoding errors over consecutive frames and the dependence of the SU data transmission on the sensing results of the previous frames. Second, the existence of other SUs in the region requires the receiver to successfully decode multiple signals from multiple sources. Thus it becomes more challenging to detect the mixed signals.

\subsubsection{Dual Radio}
A dual radio architecture is proposed in \cite{Shankar} where the receiver has one dedicated radio for sensing and one for transmission. This would improve the spectrum sensing, and more importantly, it would provide higher throughput. However, this is under the assumption that different SUs sense different channels. Otherwise, if SU1 can simultaneously sense and transmit over a channel, then SU2 would interfere with SU1 if it also starts sensing and transmitting over the same channel. In other words, quiet sensing periods are necessary if there are multiple SUs sensing the same channel even when the SU has a dual radio. Obviously, the drawback is the higher cost, higher power consumption, and higher receiver complexity.

\subsubsection{Adaptive Sensing Algorithms}
Several works have investigated the impact of using adaptive sensing time on the networks' throughput \cite{Beaulieu1, Datla, Feng1, Feng2}. Using dynamic sensing durations reduces the required number of samples, particularly when the subchannel has high SNR, and thus the overall network's throughput is significantly improved \cite{Beaulieu1}.

In \cite{Datla}, Datla et al. propose an adaptive algorithm where longer sensing durations are used for channels with high idle probabilities because they are more rewarding in terms of throughput. Each time a channel is declared to be occupied, the required sensing duration for that channel is reduced in future sensing. The drawback of this algorithm is that it can miss few available channels especially those which are declared occupied in previous sensing results. A two-phase adaptive algorithm is proposed in \cite{Feng1}.  In the first phase, the SU iteratively explores the possible set of idle channels such that after each iteration, the number of candidate channels are exponentially decreased by excluding those with low idle probability. By the second phase, the SU would have a very small subset of candidate channels, and it will allocate the sampling budget accordingly to perform fine sensing. This is extended in \cite{Feng2} such that the goal is now to detect all available idle subchannels instead of a subset of them. The algorithm significantly improves the throughput at low SNR regime. At high SNR regime, it is shown that the non-adaptive algorithm becomes optimal, where each subchannel has an equal sampling budget.  Yang et al. present a QoS-aware low complexity scheme where the SU may access some channels without spectrum sensing (i.e. $\tau=0$) \cite{Yang}. These channels have either high idle probabilities or have high tolerable interference limits. It is demonstrated that this algorithm improves the throughput. Yet, further analysis is required on how to obtain the idle probabilities of these channels and to quantify the risk of accessing a spectrum without spectrum sensing.

\subsubsection{Sequential Probability Ratio Tests (SPRTs)}
In \cite{Kim2}, parallel SPRTs are used to optimize the number of samples required for spectrum sensing (i.e. $\mathbf{o}=N$), and hence the sensing duration. Compared to fixed-sample size (FSS) detectors, parallel SPRTs significantly reduce the number of samples due to two gains: the gain of using the SPRT, and the gain of simultaneously sensing multiple bands (parallel sensing). However, a key challenge in using a bank of SPRT-branches is that each branch may yield a different sample size since, in general, the sample size is a random variable that depends on the observed data, and hence the overall sensing time would be dictated by the largest sensing delay among the parallel detectors. Caromi et al. have investigated both parallel and serial SPRTs under limited and unlimited sensing duration \cite{Caromi}. Since the SPRT has no upper limit on the number of samples required to achieve a decision, the authors propose several truncated SPRTs to limit $N$. In serial sensing, $N_{opt}$ increases as the number of sensing channels increases. On the contrary, for parallel sensing, when the number of channels increases, $N_{opt}$ decreases. In addition to that, the channel sensing order is studied in \cite{Jiang}. It is demonstrated that the intuitive sensing order, where the SU sequentially senses the channels with higher to lower idle probabilities, is not always optimum. Specifically, if non-adaptive transmission rate is used, the intuitive order is optimal, but this optimality is lost when adaptive transmission is used.

\subsubsection{Number of Cooperating Users}
Another optimization problem is
\begin{equation}
\label{eq:NumberofUsersOptimization}
\begin{aligned}
\max_{\tau,k}         \text{ }      &   C(\tau,k)                         \\
\text{s.t.}             \text{ }    &   0<k<K;~~ &   0<\tau<T,
\end{aligned}
\end{equation}
where we want to jointly optimize the number of SUs with sensing duration. Soft combining and hard combining are analyzed in \cite{Liang}. It is demonstrated that increasing the number of total SUs, $K$, reduces the optimum sensing time, and hence the throughput is improved. This is because when we have more SUs, the probability of detection increases. Thus, for a predefined target probability, we can reduce the sensing duration as we increase the number of SUs. Also, it is demonstrated that the majority-voting rule has the best performance among other hard combining rules since it provides the lowest optimum sensing time and the highest achievable throughput. Nevertheless, the gain saturates as we keep increasing the number of SUs. In \cite{Peh1}, Peh et al. investigate the optimum number of cooperating SUs, $k_{opt}$, to maximize the throughput under hard combining schemes. It is observed that the optimum value depends on the wireless environment, and hence there is no single voting rule that optimize the throughput for different SNRs. Nevertheless, optimizing $k$ reduces the sensing time and improves the throughput. Interestingly, it is demonstrated that when the channel condition is bad (low SNR), it is more advantageous to allocate more time for sensing (i.e. reduce $T-\tau$), and when it is extremely bad (extremely low SNR), then is more advantageous to allocate more time for transmission than sensing. The reason is that at such regions, $P_{FA}$ is very high regardless of $\tau$, and thus it is more beneficial to increase $T-\tau$. Soft combining is analyzed in \cite{Fan1} where it is shown that increasing the number of SUs improves the network's throughput with diminishing gain as $k$ increases.
In \cite{Mu}, soft combining and hard combining are investigated for multiband detection. It is demonstrated that soft combining provides better throughput, but hard combining causes less interference to the PUs since it has short overhead.

While increasing the number of cooperating SUs improves the reliability of detection and reduces $\tau_{opt}$, it incurs a long delay due to the time required to collect the information from all of SUs. To tackle this issue, the SUs can simultaneously send their decisions on orthogonal frequency bands \cite{Khaled1}, yet this requires larger bandwidth. Thus, another solution is sought in \cite{Wei} where the authors derive the least required number of SUs to achieve a target performance. Another technique to limit the number of cooperating SUs is censoring \cite{Axell1}. The basic principle is that the SU only cooperates if its detection result is considered useful. This technique reduces the number of cooperating users as well as save the total power budget. Similarly, selective-based cooperative spectrum sensing is presented in \cite{Alkheir1, Alkheir3, Alkheir4} where the proposed algorithms jointly reduce the overhead and mitigate the false reports sent by unreliable SUs. The basic principle is not only to limit the number of cooperating SUs, but also to admit merely those who have reliable decisions based on several factors such as the SNR of the reporting channel and the quality of sensing.

\subsection{Diversity and Sampling Tradeoffs}
Integrating the cooperative communication paradigm with multiband cognitive networks as discussed in Section IV forces a tradeoff between the spatial diversity achieved by cooperation and the expensive hardware requirements for sensing a very wideband spectrum. We quantify the latter parameter by the \emph{sampling cost}, which refers to the minimum sampling rate requirement.

\begin{figure}[!b]
\centering
\includegraphics[width=3.5in]{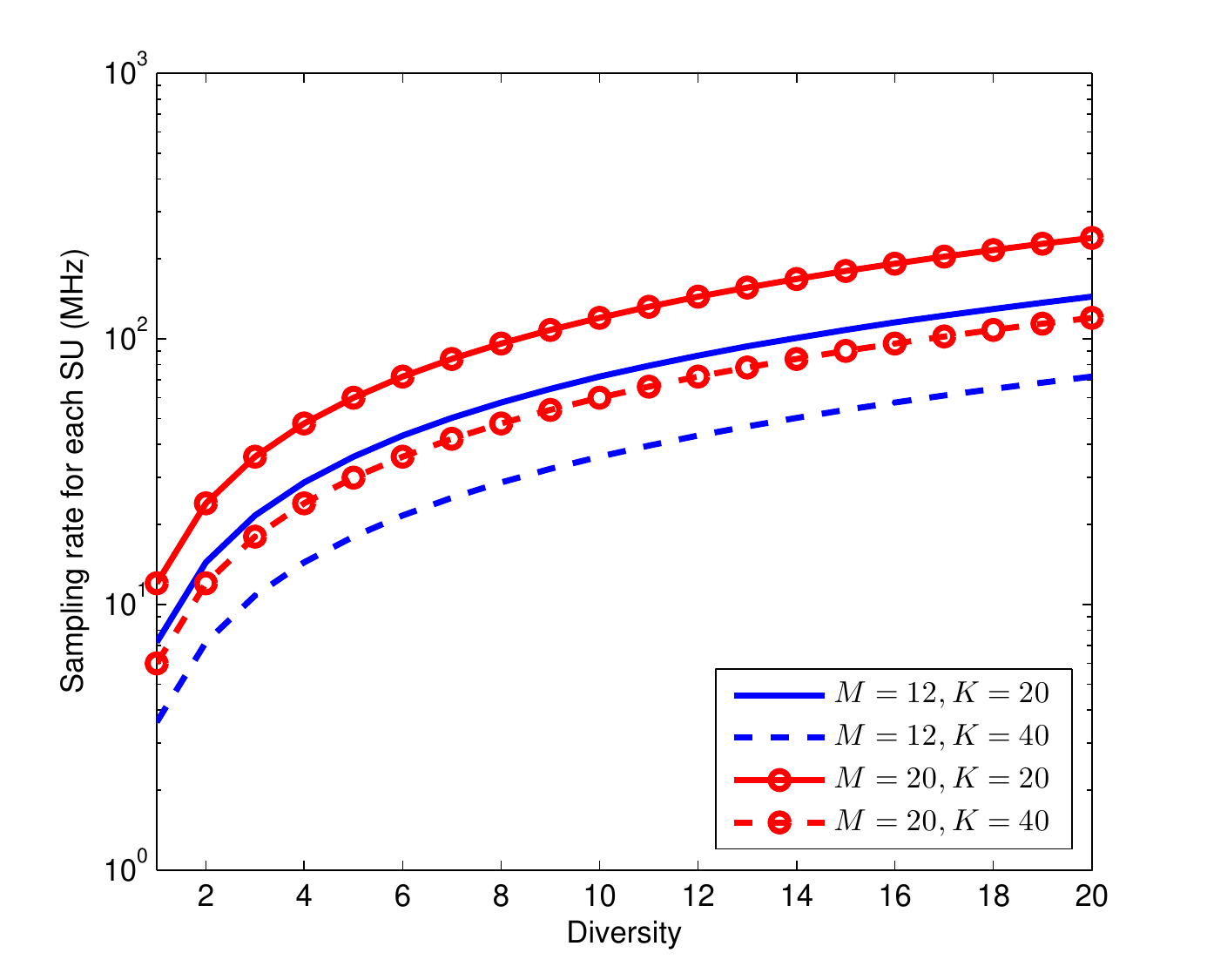}
\caption{Sampling cost-diversity tradeoff.}
\label{fig:Hatta13}
\end{figure}

\begin{figure}[!t]
\centering
\includegraphics[width=3.5in]{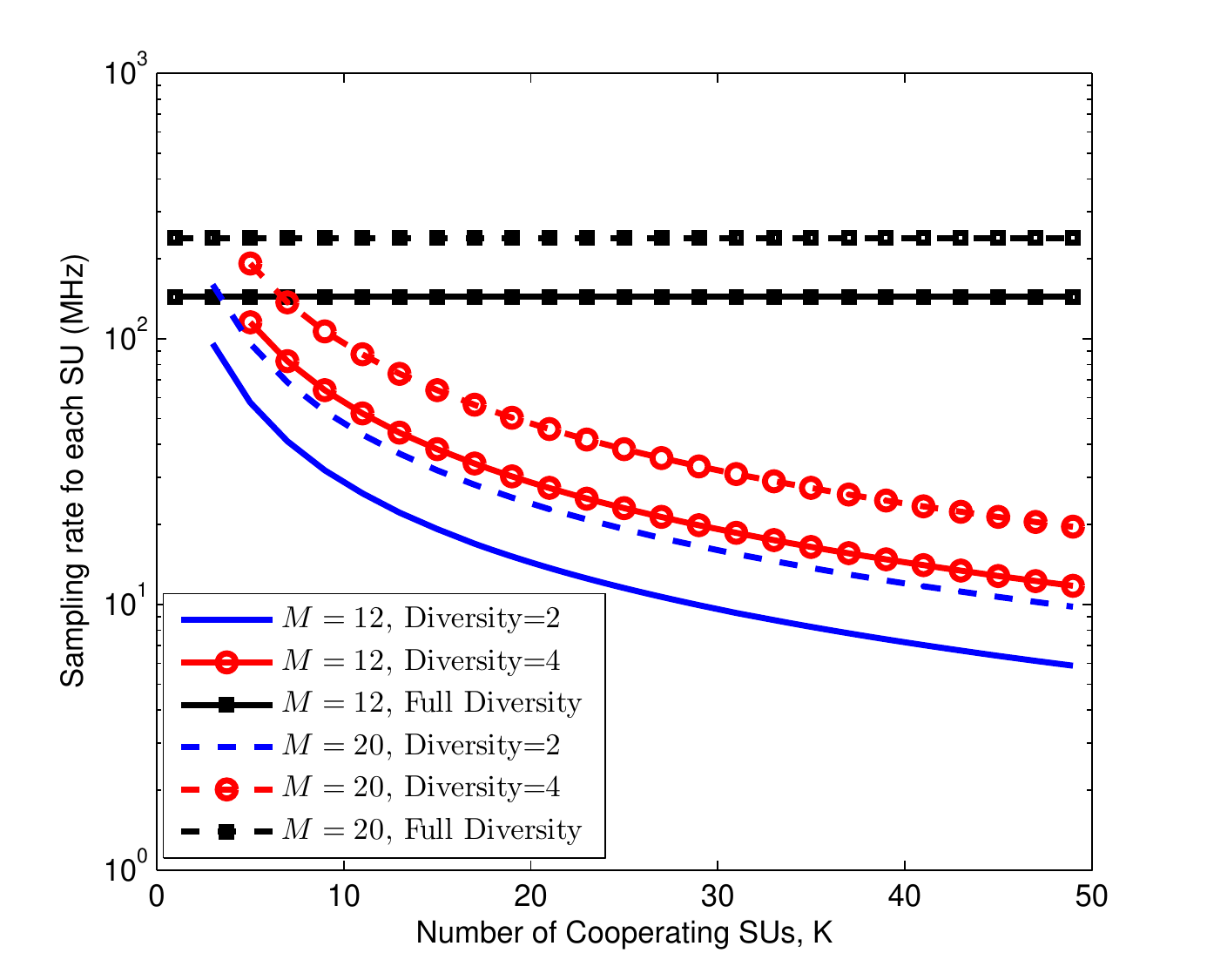}
\caption{Sampling cost versus the number of cooperating users.}
\label{fig:Hatta14}
\end{figure}

Fig. \ref{fig:Hatta13} shows that there is a tradeoff between diversity and the sampling cost. In particular, the sampling cost increases as the spatial diversity increases. It is assumed that there are $M$ subchannels with a bandwidth of 6MHz each. Also, sampling is performed at the Nyquist rate. Observe that when there are more SUs, the tradeoff's impact becomes less. This is reasonable because when we have many SUs, each SU can have a small subset of subchannels to sense (i.e. sampling cost will be reduced) for a given diversity. The impact of the number of cooperating SUs is illustrated in Fig. \ref{fig:Hatta14}. It is clear that as $K$ increases, the sampling cost decreases.  In addition, higher sampling costs are imposed when the number of subchannels, $M$, is increased. Observe that if full diversity is desired (i.e. diversity = $K$), then the sampling cost is invariant of $K$. For instance, when $M=12$, and $K=2$. Then the full-diversity is $2$, and each SU will have to monitor $12$ subchannels. Now let $K=10$. Then the full-diversity is $10$, and to achieve it, each SU must monitor 12 channels. Thus, the sampling cost is not reduced. We remark that Wang et al. show that compressive sampling can help reduce the sampling cost for a given diversity requirement by proposing a cooperative detection algorithm based on rank minimization of the SUs' collected measurement vectors \cite{Yue}.

\subsection{Power Control and Interference Limits Tradeoffs}
While power control can be categorized under resource allocation management, we dedicate a separate section for it since it has been widely studied in the literature. Optimum power allocation is vital for improving the network's throughput as well as for protecting PUs. It becomes even more important when the underlay scheme is used, since power adaptation becomes necessary. Many papers have studied joint optimization of power and sensing time to maximize the throughput. Other works include adding transmit power and interference bounds as constraint functions.

\subsubsection{Average and Peak Transmit Powers}
There are two commonly used transmit power constraints on the SUs, namely the average power, $\mathcal{P}_{avg}$, and the peak power, $\mathcal{P}_{pk}$.  The former is preferred when we want to maintain long-term power budget while the latter is used to limit the peak power for practical considerations including the non-linearity of power amplifiers \cite{Kang}. If we assume that the SUs access $l$ channels, then, the average power constraint is typically expressed as
\begin{equation}
\label{eq:AveragePower}
\sum_{m=1}^{l}  E\big[\mathcal{P}_m\big]    \leq \mathcal{P}_{avg},
\end{equation}
while the peak power constraint is expressed as
\begin{equation}
\label{eq:PeakPower}
\sum_{m=1}^{l}\mathcal{P}_m                                 \leq \mathcal{P}_{pk},
\end{equation}
To understand why the latter is more restrictive consider the following scenario, where two SUs accessing together $l$ channels, and the power bounds are  $\mathcal{P}_{avg}=\mathcal{P}_{pk}=1W$. For simplicity, assume that the transmit powers are fixed, and no fading conditions are imposed. Then during the power allocation, under the average constraint, both SUs can transmit at any power as long as the average is $1W$. A possible combination at a given time may be $\mathcal{P}_1=0.5W$, $\mathcal{P}_2=1.5W$, and so on. However, under the peak constraint, both SUs must satisfy $\mathcal{P}_1+\mathcal{P}_2\leq 1$.

Pei et al. study the joint optimization of sensing time and power allocation subject to these power constraints \cite{Pei}. It is shown that both constraints have a \emph{water-filling} power control where the SU allocates more power for the subchannels with higher SNR. The difference, however, is that for $\mathcal{P}_{pk}$, the allocation strategy depends on the number of vacant channels unlike the allocation strategy for $\mathcal{P}_{avg}$ where a certain power allocation is assigned regardless of the activity of other channels. In other words, the power allocation in $\mathcal{P}_{pk}$ is done after the SU is aware of the spectrum sensing results. It is shown that the throughput is higher when $\mathcal{P}_{avg}$  is used since it is less restrictive compared to $\mathcal{P}_{pk}$. Also, the optimum sensing time, $\tau_{opt}$, is increased when the power budget is increased. Interestingly, the variations of $\tau_{opt}$ are small for different powers, and hence using a fixed sensing time for different power budgets would only slightly degrade the throughput. Nevertheless, a fixed sensing time guarantees sensing synchronization among other SUs and reduces the network's complexity. Water-filling is also proposed by Wang in \cite{Wang} where the throughput is maximized by jointly optimizing the power allocation and the channels to be sensed. In \cite{Barbarossa}, Barbarossa et al. jointly optimize the power allocation with energy detection thresholds where it is shown that the power control is simply a multilevel water-filling procedure with levels depending on the power budget, the channel quality, and the detection reliability.

In \cite{Zou}, Zuo et al. propose an auction-based power allocation scheme when multiple SUs compete for the network resources. It works as follow. If the SU works in a spectrum interweave scheme\footnote{We refer to the access scheme where the SU only accesses a band when the PU is absent as an interweave scheme. However, the authors in \cite{Zou} refer to the same scheme as an overlay scheme, which may bring some confusion.}, the power allocated to the SU will be proportional to its payment. However, if the SU works in an underlay scheme, the power allocated is going to be constrained to protect the QoS of the PU network. An additional slot is added to the MAC frame to allow SUs bid for transmit powers as illustrated in Fig. \ref{fig:Hatta11f}. This means that for a fixed $T$, the throughput would decrease if transmission duration is decreased, or the detection performance is degraded if sensing duration is reduced instead.

In addition to that, the sensing time and power control are jointly optimized based on the distance between the SU and the PU \cite{Peh2}. The basic principle is that when the PU is far away, it is better to merely use power control to allow the SU to simultaneously exist with the PU. This is because to detect the weak PU signal, longer $\tau$ is required which impacts the throughput. However, if the PU is very close to the SU, a short sensing time would be enough to reliably detect the PU. Thus, power control is not required since it is recommended that the SU stops its transmission to protect the PU. Simulation results demonstrate that the proposed technique outperforms adaptive power allocation when the separation distance is long, and it outperforms the adaptive sensing time when the distance is short. Nevertheless, this algorithm is susceptible to inaccurate information about the PUs' locations.

\subsubsection{Average and Peak Interferences}
Interference to the PU happens when the SU simultaneously exists with the PU. This occurs when the SU either miss detects the PU, or when the SU adapts its power to meet an interference bound in the underlay scheme.
In order to protect PUs, an interference bound is set on the optimization problem. There are two common bounds, the average interference constraint $\mathcal{I}_{avg}$ and the peak interference constraint, $\mathcal{I}_{pk}$. Mathematically, the former is expressed as
\begin{equation}
 \label{eq:AverageInterference}
\sum_{i=1}^{K^{(m)}}  E[\mathcal{P}_i]  \leq \mathcal{I}_{avg},
\end{equation}
where $K^{(m)}$ is the number of SUs transmitting over subchannel $m$, and the peak interference constraint is
\begin{equation}
 \label{eq:PeakInterference}
\sum_{i=1}^{K^{(m)}}  \mathcal{P}_i  \leq \mathcal{I}_{pk}.
\end{equation}
Zhang in \cite{Zhang2} shows that the throughput of the SU network is higher when the $\mathcal{I}_{avg}$ constraint is imposed. The reason is that the average interference constraint, $\mathcal{I}_{avg}$, is generally more flexible compared to peak interference constraint, $\mathcal{I}_{pk}$, given that $\mathcal{I}_{avg}=\mathcal{I}_{pk}$. Surprisingly, while one may expect that the latter provides better protection to the PU network (since the peak bound is very restrict), it is shown otherwise due to an interesting \emph{interference diversity} phenomenon of the PU. This phenomenon is attributed to the convexity of the throughput function with respect to the interference power. It is shown that the random interference levels that arise in $\mathcal{I}_{avg}$ case are more advantageous when the interference powers are fixed as in $\mathcal{I}_{pk}$. In other words, the $\mathcal{I}_{avg}$ constraint is not only good for the SUs, but it also provides less throughput losses to the PUs! Furthermore, water-filling power allocation is  optimum under $\mathcal{I}_{avg}$, and the truncated channel inversion power allocation is optimum under $\mathcal{I}_{pk}$, which is a fair power allocation scheme that maintains a constant power by inverting the channel fading \cite{Goldsmith1}. In \cite{Hamdi}, it is recommended to impose $\mathcal{I}_{avg}$ on delay-insensitive PU systems, whereas $\mathcal{I}_{pk}$ is preferred when the systems are delay-sensitive  \cite{Hamdi} .
Stotas and Nallanathan have analyzed the impact of interference toleration to PUs on the sensing time \cite{Stotas1}. It is shown that a higher average interference bound reduces the optimum sensing time, and thus improves the throughput as expected.

\subsubsection{Beamforming}
Joint beamforming \cite{Palomar} is shown to be powerful to overcome the sensing-throughput tradeoff. Particularly, Fattah et al. demonstrate that beamforming reduces the sensing time, improves the throughput, and more importantly, it maintains a good PU protection \cite{Fattah}. Yet, it requires CSI at the SU transmitter and receiver as well as an antenna array. This is more challenging in CRNs since there are different PU networks, and the PUs may not be willing to feedback the CSI to the SUs. This motivates a new research direction on robust beamforming algorithms against imperfect CSI.

\subsubsection{Power and Resource Allocation}
In \cite{Alkheir2}, joint optimization of power control, channel selection, and rate adaptation is used to maximize the throughput. Two algorithms are proposed. In the first one, the SU selects the best available channels, and then power and rate adaptation are implemented accordingly. To guarantee that each SU picks the best channels, frequent channels reselections become inevitable, and hence high overhead is incurred. To reduce the overhead, an alternative algorithm is proposed where the SU selects a channel as long as it can support the least possible transmission rate. Otherwise, an alternative channel is randomly selected. This algorithm has lower throughput, yet it reduces the frequency of channels' reselections. This work advises to use adaptive bandwidth selection for MB-CRNs to further maximize the network's throughput.
In \cite{Panwar}, joint admission control and power allocation is studied for MB-CRNs to maximize the throughput under different constraints on QoS and power consumption. In \cite{Shi}, a joint optimization of detection thresholds, channel assignment, and power allocation is presented to maximize throughput. In \cite{Yao}, sensing time and channel selection are jointly optimized to maximize the throughput. Particularly, once the optimum sensing time for each subchannel is selected, the SU selects a subset (say $l$) from $M$ subchannels over which the aggregate throughput can be maximized. It is shown that for a fixed $M$, increasing $l$ reduces the throughput, whereas for a fixed $l$, increasing $M$ improves the throughput with a diminishing gain.

\subsection{Bandwidth Selection}
One can presume that accessing all available bands would theoretically increase the throughput. However, when a SU accesses all these bands, there is a higher probability that a PU returns to at least one of them, and thus handoff becomes necessary which consequently increases the network's overhead. Therefore, optimizing the number of subchannels for spectrum access becomes necessary. Dan et al. investigate the optimum bandwidth (or optimum number of a subset of subchannels, $l_{opt}$) to maximize network's throughput \cite{Dan}. The authors investigate both contiguous (CON) and non-contiguous (NCON) channel allocations for delay sensitive and insensitive traffic. For serial sensing, it is demonstrated that $l_{opt}$ is higher when the idle channel probability is high. Also, NCON has larger $l_{opt}$ since a larger overhead is needed to search for $l_{opt}$ contiguous channels in CON scheme. Also, CON is less sensitive to the channel idle durations compared to NCON. If the idle probability is high, it is recommended to use parallel sensing since $l_{opt}$ becomes significantly higher compared to serial sensing \cite{Dan}. If the occupancy of the channels is correlated, and the SU has prior knowledge about it, then further improvements can be attained in terms of throughput, and these benefits are observed more in CON since the contiguous channels usually have higher correlation. In addition, when there are multiple SUs in vicinity, channel reconfiguration is important in CON. To see why, imagine there are four consecutive idle bands (1 to 4) and two SUs in the network (SU1 and SU2). If $l_{opt}=2$, then if SU1 accesses bands $2$ and $3$, then SU2 cannot access $1$ and $4$ since they are non-contiguous. However, if SU1 accesses $1$ and $2$, then SU2 can be accommodated to access $3$ and $4$. Clearly, the advantage of channel reconfiguration is that it helps accommodate more SUs, but the disadvantage is the larger overhead due to the reallocation processes which incurs transmission interruptions and delays because of setting up new links. Finally, the gains of the reconfiguration scheme are shown to diminish as the number of SUs becomes larger since the losses incurred by evacuating and reconnecting to new channels overwhelm the gains. In addition to that, to enhance bandwidth efficiency, Khambekar et al. propose a novel scheme where the guard interval of OFDM symbol is utilized for spectrum sensing \cite{Khambekar}.

\section{Conclusion}
Multiband cognitive radio networks (MB-CRNs) represent the future of cognitive radio networks. This paradigm enables the secondary user to effectively utilize multiple bands simultaneously. Consequently, the network's throughput can be drastically enhanced, and thus it is expected to improve QoS provisioning . This paper has presented an in-depth analysis of the advancements of multiband spectrum sensing techniques, their challenges, and possible future directions. Moreover, cooperative networks are analyzed, and a possible extension to integrate such powerful paradigm into MB-CRN is suggested. Particularly, cooperative multiband cognitive radio provides a desirable compromise between spatial diversity and sampling complexity. In addition, some of the most common performance measures that help evaluate the network's performance in terms of spectrum reliability and network's throughput have been presented. Finally, fundamental limits and tradeoffs among several key design parameters have been thoroughly revised, and some of  the viable techniques that help reduce the impact of these tradeoffs have also been analyzed.

\bibliographystyle{IEEEtran}
\bibliography{IEEEabrv,References}

\begin{IEEEbiography}[{\includegraphics[width=1in,height=1.25in,clip,keepaspectratio]{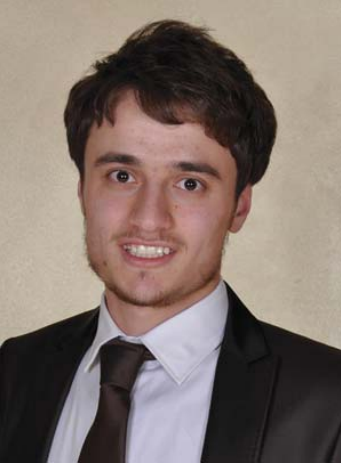}}]{Ghaith Hattab}
(S`11) received his B.Sc degree in electrical engineering from the American University of Sharjah (AUS), United Arab Emirates (UAE), in 2012, graduating \emph{summa cum laude}. He obtained his M.A.Sc in electrical and computer engineering at Queen's University, Kingston ON, Canada, in 2014.

He is the recipient of the IEEE Kingston Section Research Excellence Award in 2013, and Queen’s University Graduate Award in 2012 and 2013. He is the recipient of the prestigious Presidential Cup awarded by His Highness Sheikh Dr. Sultan Bin Mohammad Al Qassimi, member of the Supreme Council of the UAE, the Ruler of Sharjah, and President of AUS. He was also the recipient of Ministry of Presidential Affairs (MOPA) fellowship for academic excellence in the UAE.

His current research interests are in the broad area of wireless communications, and particularly, multiband cognitive radio, cooperative networks, dynamic spectrum access, MIMO communications, signal detection and estimation, and adaptive signal processing.
\end{IEEEbiography}

\begin{IEEEbiography}[{\includegraphics[width=1in,height=1.25in,clip,keepaspectratio]{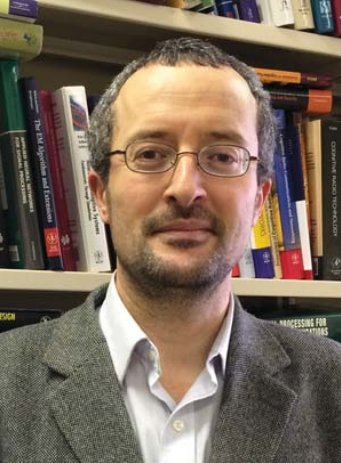}}]{Mohamed Ibnkahla}
obtained the Ph.D. degree and the \emph{Habilitation a Diriger des Recherches} degree (HDR) from the National Polytechnic Institute of Toulouse (INP), Toulouse, France, in 1996 and 1998, respectively. He obtained an Engineering degree in Electronics (1992) and a \emph{Diplome d’Etudes Approfondies} degree (equivalent to MSc) in Signal and Image Processing (1992) all from INP.

He currently holds a Full Professor position at the Department of Electrical and Computer Engineering, Queen’s University, Kingston, Canada. He has been previously an Assistant Professor at INP (1996-1999) and Queen's University (2000-2004) and Associate Professor at Queen’s University (2004-2012). He is also the founding Director of the Wireless Communications and Signal Processing Laboratory (WISIP) at Queen’s University.

His research interests include cognitive radio, adaptive signal processing, cognitive networking, neural networks, and wireless sensor networks and their applications.

He has published the following books: \emph{Signal Processing for Mobile Communications Handbook}, Taylor and Francis Publishers - CRC Press, 2004, \emph{Adaptive Signal Processing in Wireless Communications}, Taylor and Francis Publishers - CRC Press, 2008, \emph{Adaptive Networking and Cross-layer Design in Wireless Networks}, Taylor and Francis Publishers - CRC Press, 2008, \emph{Wireless Sensor Networks: A Cognitive Perspective}, Taylor and Francis Publishers - CRC Press, 2012, and \emph{Cooperative Cognitive Radio Networks: The Complete Spectrum Cycle}, Taylor and Francis Publishers - CRC Press, 2014.

He has published more than 50 peer-reviewed journal papers and book chapters, 20 technical reports, 90 conference papers, and 4 invention disclosures.

Dr. Ibnkahla received the \emph{INP Leopold Escande Medal} for the year 1997, France, for \emph{his research contributions to signal processing}; and the prestigious \emph{Prime Minister's Research Excellence Award (PREA)}, Ontario, Canada in 2001, for \emph{his contributions in wireless mobile communications.}

Dr. Ibnkahla is a Registered Professional Engineer (PEng) of the Province of Ontario, Canada.
\end{IEEEbiography}

\end{document}